\documentclass[aps,prb,showpacs,twocolumn,floats,epsfig,pdflatex]{revtex4-1}
\usepackage{amssymb}
\usepackage{amsbsy}
\usepackage{amsmath}
\usepackage{epsfig}
\usepackage{amsfonts}
\usepackage{graphicx}
\usepackage{amssymb}
\usepackage{upgreek}
\usepackage{dsfont}
\usepackage{bm}
\usepackage{color}
\usepackage{hyperref}
\usepackage{framed}
\usepackage{caption}
\usepackage{cancel}
\usepackage{subfigure}

\usepackage[title,toc,titletoc,page]{appendix}

\newcommand{\shiftleft}[2]{\makebox[0pt][r]{\makebox[#1][l]{#2}}}

\def\grad{\boldsymbol\nabla}

\begin{document}

\title {Signatures of the chiral anomaly in phonon dynamics}

\author{P. Rinkel}

\author{P. L. S. Lopes}

\author{Ion Garate}

\affiliation{D\'epartement de Physique, Institut Quantique and Regroupement Qu\'eb\'ecois sur les Mat\'eriaux de Pointe, Universit\'e de Sherbrooke, Sherbrooke, Qu\'ebec, Canada J1K 2R1}
\date{\today}
\begin{abstract}
Discovered in high-energy physics, the chiral anomaly has recently made way to materials science by virtue of Weyl semimetals (WSM). 
Thus far, the main efforts to probe the chiral anomaly in WSM have concentrated on electronic phenomena.
Here, we show that the chiral anomaly can have a large impact in the $A_1$ phonons of enantiomorphic WSM.
In these materials, the chiral anomaly produces an unusual magnetic-field-induced resonance in the effective phonon charge, which in turn leads to anomalies in the phonon dispersion, optical reflectivity, and the Raman scattering.
\end{abstract}
\maketitle

{\em Introduction.--}
A major recent development in quantum materials has been the discovery of three-dimensional Weyl semimetals (WSM)~\cite{burkov} . 
These materials contain Weyl nodes, i.e. topologically robust points of contact in momentum space between two nondegenerate and linearly-dispersing electronic bands.
Weyl nodes are characterized by a quantum number called chirality, referring to the parallel or anti-parallel locking between momentum and spin.
WSM have their low-energy electrodynamics governed by pairs of Weyl nodes with the Hamiltonian 
\begin{equation}
\label{eq:hw}
{\cal H}_W= v \tau^z {\boldsymbol\sigma}\cdot (-i {\boldsymbol\nabla}+e {\bf A} + {\bf b} \tau^z) + b_0 \tau^z - e A_0,
\end{equation}
where $v$ is the Fermi velocity, $\tau^z$ labels the two Weyl nodes of opposite chirality, $\sigma^z$ labels the two degenerate states at each node, $e$ is the electron's charge, $A^\mu=(e A_0, e v {\bf A})$ is the electromagnetic potential, and $b^\mu=(b_0,v{\bf b})$ is the axial vector describing the separation between the Weyl nodes in energy and momentum space  ($b_0$ and ${\bf b}$, respectively).

A key topological phenomenon in WSM is the chiral anomaly~\cite{fujikawa}, by which collinear electric and magnetic fields (${\bf E}$ and ${\bf B}$) induce a transfer of electrons between nodes of opposite chirality. 
This phenomenon results in an unusual electromagnetic response~\cite{burkov,sid} captured by the Lagrangian ${\cal L}_{\rm ax} = \theta {\bf E}\cdot{\bf B}$, where $\theta={\bf b}\cdot{\bf r}-b_0 t$ is a non-dynamical ``axion'' field, at position ${\bf r}$ and time $t$.
While the chiral anomaly has been probed via electronic transport experiments~\cite{huang1}, the observed signatures are not conclusive~\cite{goswami} and further probes, providing complementary understanding of the phenomenon, have been proposed~\cite{cxliu, dxiao, spivak,redell}. 
These proposals rely on the development of some type of electronic order (charge or magnetic) or nonequilibrium electronic populations.
As of now, such conditions are unmet in real WSM.

Relative to the aforementioned proposals, lattice dynamics is ubiquitous, occurs in equilibrium, and can be accurately measured.
Moreover, recent theories have predicted an interplay between lattice vibrations and electronic topology~\cite{garate1, garate2, shapourian, cortijo, pikulin}.
A natural question is then whether one might observe fingerprints of the chiral anomaly in phonon properties. 
To date, phonon measurements have found no evidence of topological effects in WSM~\cite{exp}, in part due to a scarcity of concrete theoretical ideas on what to measure. 
The objective of this work is to remedy that problem, predicting clear-cut signatures of the chiral anomaly in phonon dynamics.
Our main observation is that the coupling between Weyl fermions and phonons can produce fluctuations in $\theta$, leading to anomaly-induced infrared reflectivity, resonant Raman scattering and phonon self-energy.
We predict that such effects take place in external magnetic fields for $A_1$ phonons of \emph{enantiomorphic} WSM (materials with broken inversion and mirror symmetries), such as SrSi$_2$~\cite{huang}, trigonal Se and Te~\cite{hirayama}, Ag$_3$BO$_3$, TlTe$_2$O$_6$ and Ag$_2$Se~\cite{chang2016}.

{\em Lattice dynamics in the absence of electrons.--} 
The physical properties of lattice normal modes can be extracted from their equations of motion~\cite{keldysh},
\begin{equation}
\label{eq:phonax3}
M \left(q_0^2- \omega_{{\bf q}\lambda}^2\right) v_{{\bf q}\lambda}(q_0) = \sqrt{N} {\bf Q}^{(0)}_{-{\bf q}\lambda}\cdot{\bf E}_{\bf q}(q_0). 
\end{equation}
Here, $v_{{\bf q}\lambda}(q_0)$ is the phonon normal mode coordinate with frequency and momenta $(q_0,{\bf q})$ and branch $\lambda=1,...,3r$ for a lattice with $N$ sites and $r$ atoms in the unit cell; $M$ is the total atomic mass in a unit cell and $\omega_{{\bf q}\lambda}$ is the phonon dispersion for branch $\lambda$, in the absence of conduction electrons and photons.
The r.h.s. of Eq.~\eqref{eq:phonax3} describes a driving force, exerted by the total electric field ${\bf E}_{\bf q}(q_0)$ and proportional to the {\em mode-effective phonon charge}~\cite{keldysh,gonze1997}
\begin{equation}
\label{eq:charge0}
{\bf Q}^{(0)}_{{\bf q}\lambda} = e\sum_s  Z_s e^{-i{\bf q}\cdot{\bf t}_s}{\bf p}_{{\bf q}\lambda s}, 
\end{equation}
where $s$ labels the atoms in a unit cell, $\textbf{t}_s$ is the position of atom $s$ (measured from an origin located at the unit cell), $e Z_s$ is the effective Born charge of atom $s$ (with $\sum_s Z_s=0$), and ${\bf p}_{{\bf q}\lambda s}$ is the polarization vector describing the motion of atom $s$ in the phonon mode $\lambda$.
 In the long-wavelength limit,  Eq.~(\ref{eq:charge0}) describes the change in the intra-cell electric polarization produced by atomic displacements in mode $\lambda$. 
The modes are classified as infrared (IR) active if $Q^{(0)}_{{\bf q= 0}\lambda} \neq  0$ and IR inactive otherwise.
All acoustic modes are IR inactive due to $\sum_s Z_s =0$. 
When probing the system with visible and IR photons, we can focus on the physics of optical modes at small momenta, for which $\omega_{{\bf q}\lambda}\approx \omega_{0,\lambda}$.
As we show below, the chiral anomaly of Weyl fermions affects Eq.~(\ref{eq:phonax3}) via a dynamical renormalization of ${\bf Q}^{(0)}$.

{\em Electron-phonon interaction.--}
The departure of the atoms from their equilibrium positions induces a change $\delta U({\bf r},t)$ in the lattice potential $U({\bf r})$.
This couples locally to the electronic density as ${\cal H}_{\rm ep}=\int d^3{\bf r}\,  \psi^{\dagger}({\bf r})\psi({\bf r})\, \delta U({\bf r},t)$, where $\psi^\dagger$ is an electron creation operator. 
Focusing on electronic states in the vicinity of Weyl nodes, we have~\cite{foreman}
\begin{equation}
\label{eq:kl}
\psi({\bf r}) \simeq \frac{1}{\sqrt{\cal V}} \sum_\tau e^{i {\bf k}_\tau\cdot{\bf r}} \sum_{|{\bf k}|<\Lambda,  \sigma} e^{i {\bf k}\cdot{\bf r}} u_{\sigma\tau} ({\bf r}) c_{{\bf k} \sigma\tau},
\end{equation}
where ${\cal V}$ is the sample volume,  ${\bf k}_\tau$ is the position of node $\tau$ in momentum space, $u_{\sigma\tau}({\bf r}) \equiv u_{{\bf k}_\tau \sigma}({\bf r})$ is the periodic part of the Bloch wave function at node $\tau$, ${\bf k}$ is the momentum measured from a node, and $\Lambda$ is a high-momentum cutoff (smaller than the internode separation in ${\bf k}$ space). 
Hereafter, we will be interested in long wavelength phonons ($|{\bf q}|\ll|{\bf k}_\tau-{\bf k}_{\tau'}|$ for $\tau\neq \tau'$), which are unable to scatter electrons between different nodes.
In this case, the electron-phonon interaction Hamiltonian reads 
\begin{equation}
\label{eq:hep0}
{\cal H}_{\rm ep} = \sum_{{\bf k}{\bf q}} \sum_{\sigma\sigma'\tau} \left[\sum_\lambda  g^\lambda_{\sigma\sigma',\tau}({\bf q}) v_{{\bf q}\lambda}(t)\right] c^\dagger_{{\bf k}\sigma\tau} c_{{\bf k}-{\bf q}\sigma'\tau},
\end{equation}
where $|{\bf k}|,|{\bf k}-{\bf q}|<\Lambda$ and
\begin{equation}
\label{eq:gfull} 
g^\lambda_{\sigma\sigma',\tau}({\bf q})= \frac{\sqrt{N}}{{\cal V}}\sum_s  {\bf p}_{{\bf q}\lambda s} \cdot \langle u_{\sigma\tau}|e^{-i\mathbf{q}\cdot{\mathbf{r}}}\frac{\partial U\left({\mathbf{r}}-\mathbf{t}_{s}\right)}{\partial\mathbf{t}_{s}}|u_{\sigma'\tau}\rangle
\end{equation}
is the electron-phonon coupling.
It is useful to decompose Eq.~(\ref{eq:gfull}) as
\begin{equation}
\label{eq:deco}
g^\lambda_{\sigma\sigma',\tau} = g^\lambda_{0 0}\delta_{\sigma\sigma'} + {\bf g}^\lambda_0 \cdot{\boldsymbol\sigma}_{\sigma\sigma'} + \tau (g^\lambda_{0 z}\delta_{\sigma\sigma'} + {\bf g}^\lambda_z\cdot{\boldsymbol\sigma}_{\sigma\sigma'}),
\end{equation}
where ${\boldsymbol\sigma}$ is a vector of Pauli matrices and $\tau=\pm 1$.
The expansion coefficients are explicitly listed in the appendix~\ref{app:appB}.
Comparing Eqs.~(\ref{eq:hep0}) and (\ref{eq:deco}) with Eq.~(\ref{eq:hw}), we infer that
$c^{\mu}({\bf q},t)\equiv\sum_\lambda\left(g_{00}^{\lambda},\mathbf{g}_{z}^{\lambda}\right)v_{\mathbf{q}\lambda}\left(t\right)$
and
$c_5^{\mu}({\bf q},t)\equiv\sum_\lambda \left(g_{0z}^{\lambda},\mathbf{g}_0^{\lambda}\right)v_{\mathbf{q}\lambda}\left(t\right)$
behave as effective electromagnetic and axial-vector fields acting on electrons (respectively).
In particular, $c_5^\mu ({\bf q}\simeq 0, t)$ can be interpreted as phonon-induced fluctuations in the energy difference and momentum separation between Weyl nodes of opposite chirality.

While $g^\lambda_{00}$ is generally allowed by symmetry, $g^\lambda_{0 z}\neq 0$ requires $\sum_\tau |u_{\sigma\tau}({\bf r})|^2 \tau \neq 0$, which in turn demands a WSM belonging to an enantiomorphic point group.
Similarly, ${\bf g}^\lambda_z\neq 0$ requires broken time-reversal (TR), inversion and mirror symmetries, while ${\bf g}^\lambda_0\neq 0$ requires breaking of TR symmetry. 
In enantiomorphic WSM with TR symmetry, pairs of nodes related by TR give additive contributions to $g_{0z}^\lambda$. 


{\em Phonon dynamics in presence of Weyl fermions.--}
Weyl fermions modify the phonon equations of motion by means of an effective action 
\begin{equation}
\label{eq:ds}
S_{\rm int} = -i\ln\,{\rm det}\left(i\gamma^\mu \partial_\mu - \gamma^\mu a_\mu - \gamma^5 \gamma^\mu a_{5\mu}\right).
\end{equation}
Here, $\gamma^\mu=(\tau^x,i\tau^y{\boldsymbol\sigma})$ and $\gamma^5=\tau^z$ are $4\times 4$ Dirac matrices, whereas
$a_{5\mu}=b_\mu +c_{5\mu}$ and $a_\mu=A_\mu+c_\mu$ are the total axial-vector and vector fields including the phonon parts.
The inclusion of electromagnetic fields in Eq.~(\ref{eq:ds}) is essential because we are interested in optical effects produced by phonons.

To evaluate the influence of Weyl fermions in the phonon dynamics, we expand
$S_{\rm int } = S_2+S_3+...$ in powers of $a^\mu$ and $a_5^{\mu}$.
Afterwards, we solve for $\delta S/\delta v_{{\bf q}\lambda}(q_0) =0$, where $S=S^{(0)}+S_{\rm int}$ is the effective action for phonons and $S^{(0)}$ is the bare phonon action (the latter of which yields Eq.~(\ref{eq:phonax3})).
Explicitly,
\begin{align}
\label{eq:delS}
S_2 &= \int_q \Pi_{\mu\nu}(q) \left[a^\mu(q) a^\nu(-q) + a_5^\mu(q) a_5^\nu(-q)\right] \nonumber\\
S_3 &=\int_{k,k'} T_{\alpha\mu\nu}(k,k') a^\mu(k) a^\nu (k') a_5^\alpha(-k-k'),
\end{align}
where $\int_k \equiv \int d^4 k /(2\pi)^4$ and $k^\mu=(k_0, v {\bf k})$.
The absence of terms with an odd number of $a^\mu$ is a consequence of assuming zero chemical potential (charge conjugation symmetry).
In addition, we take zero temperature and keep terms that are up to second order in $v_{{\bf q}\lambda}$ (harmonic approximation).

The effective Lorentz symmetry of Eq.~(\ref{eq:hw}), together with gauge invariance,  fixes $\Pi_{\mu\nu} (q)= (q_\mu q_\nu - g_{\mu\nu} q^2) \Pi(q^2) $, where $q^2=q_0^2-v^2|{\bf q}|^2$ and  $\Pi(q^2)$ is the polarization function~\cite{peskin}. 
Although $\Pi(q^2)$ diverges as $1/{\rm log}(q^2)$~\cite{dxiao,zhang} at $q^2=0$ ($q_0= \pm v |{\bf q}|$), we find that $S_2$ does not contribute qualitatively to the phonon dynamics, unlike the contribution from the chiral anomaly (see appendix~\ref{app:appC}). 

The chiral anomaly emerges from $S_3$, through the amplitude $T_{\mu\nu\lambda}$ associated to the ``VVA triangle diagram''~\cite{adler}.
This amplitude has been thoroughly studied in high-energy physics~\cite{bertlmann}, though one important caveat in our case is that Weyl fermions and photons have different phase velocities ($v$ and $c$).
Consequently, $k^2= (c^2-v^2)|{\bf k}|^2 \neq 0$ for real photons.
It is helpful to decompose $T_{\alpha\mu\nu}$ into longitudinal and transverse parts~\cite{armallis2009},
\begin{equation}
\label{eq:tri}
T_{\alpha\mu\nu}(k,k')=T^{(l)}_{\alpha\mu\nu}(k,k')+ T^{(t)}_{\alpha\mu\nu} (k,k'),
\end{equation}
such that $q^\alpha T^{(l)}_{\alpha\mu\nu}\neq 0$ and $q^\alpha T^{(t)}_{\alpha\mu\nu}=0$, where $q=k+k'$.
The Ward identities $k^\mu T_{\alpha\mu\nu}=k'^\nu T_{\alpha\mu\nu}=0$ ensure the gauge invariance of $S_3$.
In particular, the longitudinal part reads
\begin{equation}
T^{(l)}_{\alpha\mu\nu}(k,k')=w_L(q^2) q_\alpha\epsilon_{\mu\nu\rho\sigma} k^\rho k'^\sigma+\mathcal{O}((k^2+k'^2)/\Lambda^2),
\end{equation}
where $\epsilon_{\mu\nu\rho\sigma}$ is the Levi-Civita tensor and $w_L = - i/(2 \pi^2 q^2)$. 
The \emph{infrared} ($1/q^2$) pole in $w_L$ is the essence of the chiral anomaly. 
It is independent of the existence of high-energy bands, as well as of the cutoff $\Lambda$; it depends on material properties only through the Fermi velocity $v$ hidden in $q^2$. 
Also, the chiral anomaly survives under perturbations that break Lorentz invariance (such as nonzero temperature and chemical potential, or disorder), although the $q^2=0$ pole gets broadened into a resonance~\cite{itoyama}.

The transverse amplitude $T^{(t)}$, more complicated, is unrelated to the the chiral anomaly, but does contribute to the phonon dynamics and can cancel the $1/q^2$ pole of $T^{(l)}$ in particular kinematic conditions (as when either $k^2$ or $k'^2$ are nonzero~\cite{armallis2009}). 

Nevertheless, the analysis becomes simple when 
the magnetic field contains a constant and uniform part ${\bf B}_0$ (see appendix~\ref{app:appC}).
In this case, $T^{(t)}$ is non-singular and may be neglected, while $T^{(l)}$ changes Eq.~(\ref{eq:phonax3}) via
${\bf E}_{\bf q}\to {\bf E}_{\bf q}(q_0)+{\bf E}_{\bf q}^{\rm ph}$ and ${\bf Q}^{(0)}_{{\bf q}\lambda} \to {\bf Q}^{(0)}_{{\bf q}\lambda}+\delta{\bf Q}_{{\bf q}\lambda}(q_0)$.
Here, 
${\bf E}_{\bf q}^{\rm ph}(q_0) = (i/e v) \sum_\lambda \left(v {\bf q} g_{00} - q_0 {\bf g}_z\right) v_{{\bf q}\lambda}$ is an effective electric field related to phonons, whereas
\begin{equation}
\label{eq:qp}
\delta{\bf Q}_{{-\bf q}\lambda}(-q_0) =i\frac{e^2 {\cal V}}{\pi^2 \hbar^2 \sqrt{N}} \frac{{\bf B}_0}{q^2} (q_0 g^\lambda_{0z}-v{\bf q}\cdot{\bf g}^\lambda_0)
\end{equation}
is a phonon effective charge induced by a magnetic field and mediated by Weyl fermions.
Equation ~(\ref{eq:qp}) is the main finding of this work. 
It can be reinterpreted as a phonon-modulated topological magnetoelectric polarization: $\delta{\bf Q}_{{\bf q}\lambda} = \partial^2 {\cal L}_{\rm ax}/(\partial v_{{\bf q}\lambda}\partial {\bf E}_{-\bf q})$, where ${\cal L}_{\rm ax}=(\theta+ \delta\theta_{\bf q}) {\bf E}_{-\bf q}\cdot{\bf B}_0$ and
$\delta\theta_{\bf q}\propto i \sum_\lambda(q_0 g^\lambda_{0 z} - v{\bf q}\cdot{\bf g}^\lambda_0)v_{{\bf q}\lambda} /q^2$ is a phonon-induced axion. 

Magnetically induced effective phonon charges are not unique to WSM: they also occur e.g. in multiferroic materials~\cite{vermette}.
Indeed, according to group theory, phonons that are IR inactive at ${\bf B}_0=0$ may become IR active at ${\bf B}_0\neq 0$, provided that they belong to the direct product of axial and polar irreducible representations (irreps)~\cite{anasta}.
What is unique about WSM is the microscopic  mechanism for the ${\bf B}$-induced IR activity, namely that $\delta{\bf Q}$ originates from axial electron-phonon interactions $(g^\lambda_{0z}, {\bf g}^\lambda_0)$ and that it contains a chiral-anomaly-induced \emph{pole} at $q_0=\pm v|{\bf q}|$.
It is crucial that the ${\bf B}$-field have a nonzero static and uniform component; otherwise the pole would get cancelled by $T^{(t)}$.

Because $\theta$ is a pseudoscalar, $\delta\theta_{\bf q}=\sum_\lambda(\partial\theta/\partial v_{{\bf q}\lambda}) v_{{\bf q}\lambda}$ can be nonzero only if at least one of the modes is pseudoscalar.
Pseudoscalar phonons transform as ${\bf E}\cdot{\bf B}$ under proper and improper rotations.
Often, Weyl nodes are located at arbitrary points in the Brillouin zone, where the Bloch states transform according to a one dimensional irrep of the translation subgroup. 
In such case~\cite{powell}, only phonons that transform as the totally symmetric irrep ($A_1$) couple to Weyl fermions in the long wavelength limit.
Moreover, for $A_1$ phonons to be pseudoscalar, the crystal must lack mirror planes~\cite{anasta}.
Hence, we predict that the main effects of the chiral anomaly will manifest themselves in $A_1$ phonons of enantiomorphic WSM.




In order to assess the observability of our predictions, we estimate $\delta Q$.
In a TR-symmetric (${\bf b}={\bf g}_0^\lambda={\bf g}_z^\lambda=0$) and enantiomorphic ($b_0\neq 0 \neq g_{0 z}^\lambda$) WSM, we have
\begin{equation}
\label{eq:guesst}
\frac{|\delta Q_{\bf q}(q_0)|}{e}\sim \frac{I_0}{\hbar q_0} \frac{q_0^2}{q_0^2-v^2 {\bf q}^2} \frac{{|\bf B}_0| {\cal V}_c/a_B}{\phi_0} \frac{b_0}{W},
\end{equation}
where $a_B$ is the Bohr radius, $I_0$ is a Rydberg, ${\cal V}_c$ is the unit cell volume, $\phi_0$ is the quantum of flux and $W\sim O(\Lambda)$ is a characteristic electronic bandwith. 
In this estimate, we have assumed that the spatial range of $\partial_{{\bf t}_s} U({\bf r}-{\bf t}_s)$ is about a unit cell, that its magnitude within that range is about $I_0/a_B$, and that $\sum_\tau \tau |u_{\sigma\tau}|^2 \simeq (b_0/W) \sum_\tau |u_{\sigma\tau}|^2$.
For ${|\bf B}_0|\sim 1 {\rm T}$, ${\cal V}_c \sim 125 \AA^3$, $I_0/(\hbar q_0)\sim 10^3$ and $b_0/W\sim 0.1$, we have 
$|\delta Q_{\bf q}(q_0)|\sim 0.1 e (1- v^2 |{\bf q}|^2/q_0^2)^{-1}$,
which is not negligible (especially close to resonance). 
The unusual frequency- and momentum-dependence of $\delta {\bf Q}$ leads to new physical effects that we discuss next.

{\em Phonon dispersion.--}
In the electrostatic approximation~\cite{stroscio}, valid for $c |{\bf q}| \gg q_0$, phonons produce longitudinal electric fields ${\bf E}_{\bf q}(q_0)\simeq -(\sqrt{N}/\epsilon_e {\cal V}) \hat{\bf q}\sum_\lambda ({\bf Q}_{{\bf q}\lambda}\cdot\hat{\bf q})\, v_{{\bf q}\lambda}(q_0)$, where $\epsilon_e ({\bf q},q_0)$ is the electronic dielectric function for Weyl fermions (see appendix~\ref{app:appD}). 

For simplicity we consider a time-reversal invariant WSM and assume low carrier concentrations ($\omega_{\rm plasma}\ll\omega_{{\bf q}\lambda}$), so that the plasmon-phonon hybridization is unimportant. 
Also, as an illustration, we consider a single phonon mode: we assume that it is IR inactive at zero magnetic field, but that it couples axially to electrons.
The $A_1$ phonons in SrSi$_2$, Ag$_2$Se and CoSi could be candidates for such a mode~\cite{anasta,bilbao}.

\begin{figure}[t]
    \centering\hspace{-2mm}
    \captionsetup{justification=raggedright,margin=0cm}
     \includegraphics[width=1\columnwidth]{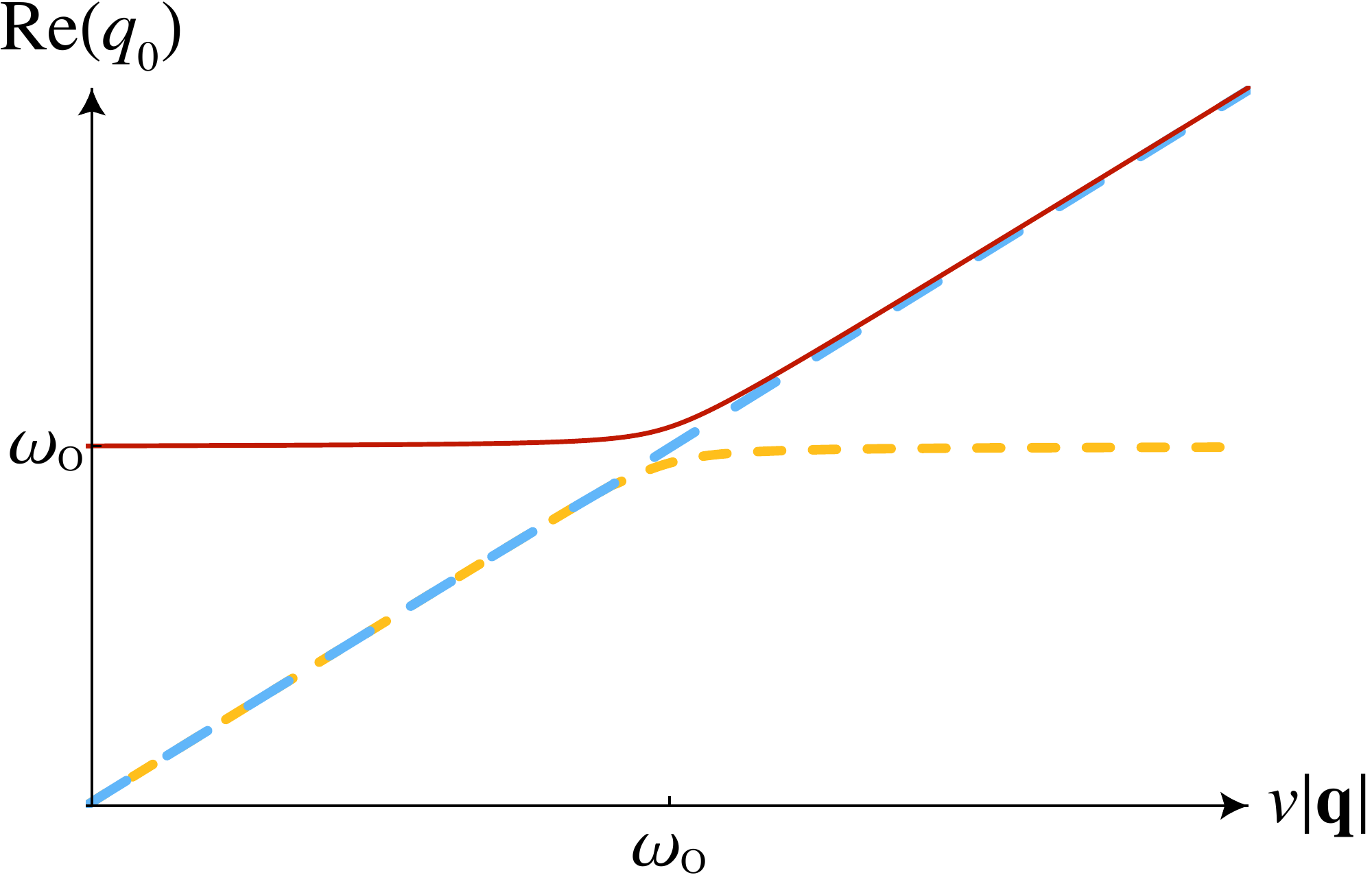}\llap{ \put(-120,0){\raisebox{0.7cm}{\fbox{\includegraphics[width=0.32\columnwidth]{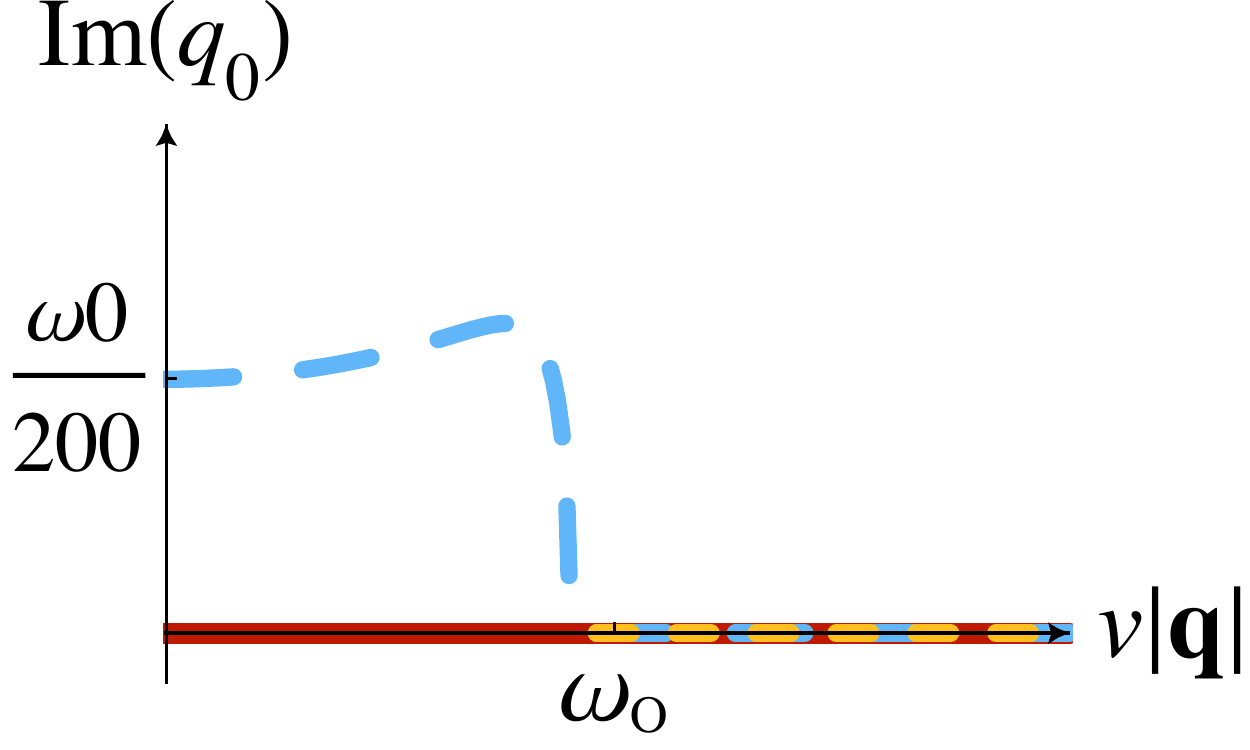}}}}} 
\caption{Anomaly-induced coupling between an IR inactive optical phonon and a linearly dispersing pseudoscalar boson in presence of a static and uniform magnetic field ${\bf B}_0$, with $\hat{\bf q}\cdot{\bf B}_0 = -1 T$ (see appendix~\ref{app:appD} for plots with other values of $\hat{\bf q}\cdot{\bf B}_0$).
The curves are solutions to Eq.~(\ref{eq:lo}), with parameter values taken from appendix~\ref{app:appD}. When $|{\bf q}|\lesssim\omega_0/v$, the linear mode is doubly degenerate but one solution is unstable (anti-damped). A large static dielectric constant is assumed so that Landau damping at $q_0>v|{\bf q}|$ can be neglected.}
\label{fig:pho}
\end{figure}

Inserting ${\bf E}$,  ${\bf E}^{\rm ph}$ and $\delta{\bf Q}$ in Eq.~(\ref{eq:phonax3}), we obtain the phonon dispersion from
(see appendix~\ref{app:appD})
\begin{equation}
\label{eq:lo}
\omega_{{\bf q} A_1}^2-q_0^2 + i \kappa g_{00}^{A_1} {\bf q}\cdot \delta{\bf Q}_{{\bf q} A_1}(q_0) + \eta\frac{|\hat{\bf q}\cdot\delta{\bf Q}_{{\bf q} A_1} (q_0)|^2}{\epsilon_e} =0,
\end{equation}
where $\kappa= \sqrt{N}/(M e)$ and $\eta=N/(M {\cal V})$.
The solution of this equation is displayed in Fig.~\ref{fig:pho}. 
When ${\bf B}_0\cdot\hat{\bf q} =0$, the only solution is $q_0=\omega_{{\bf q} A_1}\simeq\omega_0$ (for long wavelength phonons).
As soon as ${\bf B}_0\cdot\hat{\bf q}\neq 0$, a new mode appears due to the anomaly pole, which has quasilinear dispersion.
This mode describes particle-hole pairs propagating at the Fermi velocity, and is the analogue of the pseudoscalar boson discussed in high-energy physics~\cite{gianotti,itoyama}. 
Remarkably, the linear mode couples to the optical phonon in the vicinity of $q_0\simeq v |{\bf q}|$, somewhat like ordinary photons and optical phonons couple in the vicinity of $q_0\simeq c |{\bf q}|$.
When $g_{00}^{A_1}= 0$, the gap between the optical phonon and the pseudoscalar boson at $|{\bf q}|\simeq \omega_0/v$ scales as $|\hat{\bf q}\cdot{\bf B}_0|^{2/3}$.

{\em Raman scattering.--}
One-phonon Raman scattering arises from first order corrections to the electronic susceptibility by lattice displacements~\cite{cardona1982}. 
The amplitude of Raman scattering can be represented by a triangle diagram with two photon lines and a phonon line. 

Let us consider the case where no static magnetic fields are present.
In this case, the Raman scattering of a pseudoscalar $A_1$ phonon is directly linked to Eq.~(\ref{eq:tri}) and its contribution to the Raman tensor contains an ``axial'' component 
\begin{equation}
\label{eq:anram}
R^{\rm ax}_{j j' A_1}\propto T_{\alpha\mu\nu}(k,k') \frac{\partial^3\left[ c_5^\alpha(q) A^\mu (k) A^\nu (k')\right]}{\partial {\bf E}_j(k) \partial {\bf E}_{j'} (k') \partial v_{{\bf q}A_1}(q_0)},
\end{equation}
where $j,j'\in\{x,y,z\}$ denote the polarizations of the incoming and scattered electric fields ${\bf E}(k)$ and ${\bf E}(k')$, respectively, 
$k=-(\omega, v {\bf k})$ and $k'=(\omega',v {\bf k}')$ are the momenta of incoming and scattered photons, and $q=k+k'=(q_0, v {\bf q})$ is the phonon frequency and momentum.
Equation~(\ref{eq:anram}) describes the contribution to the Raman tensor coming from phonon modulations of the {\em magnetoelectric} polarizability.
Because $k$ and $k'$ are in the visible~\cite{cutoff}, 
the $1/q^2$ pole in $T^{(l)}$ is cancelled by $T^{(t)}$.
Yet, a weaker singularity remains near $q_0=v|{\bf q}|$ (see appendix~\ref{app:appE}), 
\begin{align}
\label{eq:anram2}
&R^{\rm ax}_{j j' A_1}\Big|_{q^2\simeq 0} \propto
\frac{\left(q_0 g_{0 z} - v {\bf q}\cdot{\bf g}_0\right)}{(k^2-k'^2)^3}\nonumber\\
&\times\left[k^4 \ln\left(\frac{k^2}{q^2}\right) - k'^4 \ln\left(\frac{k'^2}{q^2}\right)\right] \epsilon_{j j' l} \left(\hat{\bf k}'-\hat{\bf k}\right)_l,
\end{align}
where $l\in\{x,y,z\}$ and $\epsilon_{j j' l}$ is the Levi-Civita tensor.
Aside from being antisymmetric under $j\leftrightarrow j'$, 
 the $\ln(q^2)$ singularity in $R^{\rm ax}_{j j' \lambda}$ 
is independent from the ultraviolet cutoff of the theory, i.e. associated to low-energy universal properties of 3D Dirac fermions.
This anomaly of the Raman tensor appears experimentally accessible for typical optical phonons, because the momentum $q_0/v\simeq 5\times 10^5 {\rm cm}^{-1}$ is achievable in the backscattering configuration.

{\em Infrared reflectivity.--}
Phonon modes with nonzero mode-effective charge produce fluctuating dipole moments that couple to electromagnetic fields. 
This coupling is quantified by the lattice dielectric susceptibility~\cite{keldysh}.
Like above, let us consider the case of an $A_1$ phonon that is IR inactive at zero magnetic field, and couples to electrons axially. 
This mode's contribution to the lattice susceptibility reads (see appendix~\ref{app:appD})
\begin{equation}
\label{eq:sus}
\chi_{jj^{'}}^{\rm latt}\left(\mathbf{q},q_{0}\right)=\frac{1}{M\mathcal{V}_{c}} \frac{\delta Q_{{\bf q}\lambda j}\delta Q_{{\bf q}\lambda j'}}{\omega_{{\bf q}\lambda}^2 + i \kappa g_{00}^\lambda {\bf q}\cdot \delta{\bf Q}_{{\bf q}\lambda}-q_0^2},
\end{equation}
where $j,j'\in\{x,y,z\}$.
Thus, a constant and uniform magnetic field will induce an IR absorption in an otherwise IR inactive mode.
In addition, the absorption spectrum depends on $\hat{\bf q}\cdot{\bf B}_0$. 
This effect can be probed in optical reflectivity experiments~\cite{wooten}, e.g. in the following configurations: (1) non-normal incidence of light whose polarization is not parallel to the sample surface, with ${\bf B}_0$ along the normal to the surface; (2) normal incidence of light, whose polarization is parallel to the sample surface, with ${\bf B}_0$ parallel to the sample surface.
In optical experiments  $q_0 = c |{\bf q}|$ is fixed and hence the resonance of $\delta Q$ at $q_0 = v |{\bf q}|$ is out of reach.
Alternative probes (inelastic X-ray scattering, electron energy loss spectroscopy) may allow to access the most interesting regime ($q_0 \simeq \omega_{{\bf q}\lambda}$ and $|{\bf q}| \simeq \omega_{{\bf q}\lambda}/v$).

In conclusion, we have predicted a resonant magnetic-field-induced phonon charge as a new fingerprint of the chiral anomaly.
This translates into a resonant Raman scattering, a magnetic-field-induced infrared activity, and a peculiar magnetic-field-dependence of the dispersion of $A_1$ phonons in enantiomorphic Weyl semimetals. 
Although our main results involve optical phonons, anomaly-induced effects may be present in the acoustic phonons as well.
In addition, the dynamical screening of the electron-phonon interactions, not mentioned above, does not change our results substantially.
Further analysis of these issues will be subject of future work.

{\em Acknowledgements.--}
IG acknowledges the hospitality of the Spin Phenomena Interdisciplinary Center (SPICE), where this work was initiated.
We have benefitted from fruitful discussion with K. Burch, A. Grushin, D. Pesin, B. Roberge and S. Xu. 
PR and IG are funded by Qu\'ebec's RQMP and Canada's NSERC.
PLSL is supported by the Canada First Research Excellence Fund. 

{\em Note added.--} 
Recently, we noticed work that overlaps with
some of our results \cite{song}.

\begin{widetext}

\appendix

\section{Lattice dynamics in absence of electrons \label{app:appA}}

In this section, we review for pedagogical purposes the derivation of the equations of motion for lattice vibrations in a system without electron-phonon coupling.
We follow closely the treatment and conventions of Ref.~\cite{keldysh}.

Let the displacement of atom $s$ in unit cell ${\bf l}$ at time $t$ away from its equilibrium position be  denoted as ${\bf w}_{{\bf l}s}(t)$.
In the harmonic approximation, the Hamiltonian of the lattice is
\begin{equation}
{\cal H}_{\rm latt} = \sum_{{\bf l} s j } \frac{p_{{\bf l}s j}^2}{2 M_s} + \frac{1}{2} \sum_{{\bf l} s j }\sum_{{\bf l}'s' j} A_{s j, s' j'}({\bf l}-{\bf l}') w_{{\bf l} s j} w_{{\bf l}' s' j'},
\end{equation}
where $M_s$ is the mass of atom $s$, ${\bf p}_{{\bf l}s}$ is the momentum of the atom $s$ in unit cell ${\bf l}$, $A_{s j, s' j'}({\bf l}-{\bf l}')$ is the force constants matrix and $j,j'\in\{x,y,z\}$.
Thus, the equations of motion for the displacements are
\begin{equation}
\label{eq:coup}
M_s \ddot{w}_{{\bf l} s j} + \sum_{{\bf l}' s' j'} A_{s j,s'j'}({\bf l}-{\bf l}') w_{{\bf l}' s' j'} =0.
\end{equation}
The solutions for these equations can be casted as
\begin{equation}
\label{eq:ws}
{\bf w}_{{\bf l}s} = {\bf w}_s e^{i({\bf q}\cdot{\bf l} -\omega_{\bf q} t)},
\end{equation}
where $w_{s j}$ may be viewed as the $3 r$ components of a vector, $r$ being the number of atoms in the unit cell.
Replacing Eq.~(\ref{eq:ws}) in Eq.~(\ref{eq:coup}), we have
\begin{equation}
\label{eq:eigen}
-M_s \omega_{{\bf q}}^2 w_{s j} +\sum_{s'j'} A_{s j, s' j'}({\bf q}) w_{s'j'} =0,
\end{equation}
where
\begin{equation}
A_{s j, s' j'}({\bf q}) = \sum_{\bf l} A_{s j, s' j'} ({\bf l}) e^{-i {\bf q}\cdot{\bf l}}.
\end{equation}
The solution of Eq.~(\ref{eq:eigen}) is nontrivial if
\begin{equation}
||\, M_s \omega_{\bf q}^2 \delta_{s s'} \delta_{j j'} - A_{s j, s' j'}({\bf q}) \,|| = 0.
\end{equation}
This is an equation of degree $3 r$.
We denote its roots by $\omega_{{\bf q}\lambda}$, where $\lambda=1,..., 3 r$ is the normal mode index.
If the normal modes take into account the short-range electric fields caused by the ionic motion,  $\omega_{{\bf q}\lambda}$ can have a nonzero imaginary part.
Corresponding to each frequency $\omega_{{\bf q}\lambda}$, there is a polarization vector ${\bf p}_{{\bf q}\lambda s}$ that satisfies Eq.~(\ref{eq:eigen}).
The vectors ${\bf p}_{{\bf q}\lambda s}$ constitute a set of $3 r$ numbers that give the ratio of displacements ${\bf w}_s$ of atoms.
Each frequency $\omega_{{\bf q}\lambda}$ and polarization vector ${\bf p}_{{\bf q}\lambda s}$ corresponds to one irreducible representation of the symmetry group of the crystal.
The polarization vectors satisfy the orthogonality condition
\begin{equation}
\sum_s M_s {\bf p}^*_{{\bf q}\lambda s} \cdot{\bf p}_{{\bf q}\lambda' s} = M \delta_{\lambda\lambda'},
\end{equation}
where $M=\sum_s M_s$ is the mass of all atoms in the unit cell.
Likewise,
\begin{equation}
\sum_{{\bf l}s} e^{i({\bf q}-{\bf q}')\cdot{\bf l}} M_s {\bf p}^*_{{\bf q}\lambda s} \cdot{\bf p}_{{\bf q}'\lambda' s}  =  N M \delta_{{\bf q} {\bf q}'}\delta_{\lambda\lambda'},
\end{equation}
where $N$ is the number of unit cells in the crystal.
Thus,
\begin{equation}
\frac{1}{\sqrt{N}} e^{i{\bf q}\cdot{\bf l}} {\bf p}_{{\bf q}\lambda s}
\end{equation}
forms a complete set of vectors that can be used to expand any displacement vector:
\begin{equation}
\label{eq:wl}
{\bf w}_{{\bf l}s} = \frac{1}{\sqrt{N}} \sum_{{\bf q} \lambda} e^{i{\bf q}\cdot{\bf l}}\, {\bf p}_{{\bf q}\lambda s} v_{{\bf q}\lambda} \equiv  \sum_{{\bf q}} e^{i{\bf q}\cdot{\bf l}} {\bf w}_{{\bf q}s}.
\end{equation}
The coefficient of expansion  $v_{{\bf q}\lambda}$ is the normal mode coordinate of the lattice vibrations:
\begin{equation}
v_{{\bf q}\lambda}=\frac{1}{\sqrt{N}}\sum_{{\bf l} s}  e^{-i{\bf q}\cdot {\bf l}}\,\frac{M_s}{M} {\bf w}_{{\bf l}s}\cdot{\bf p}^*_{{\bf q}\lambda s}.
\end{equation}
In the second quantized formalism, $v_{{\bf q}\lambda}$ becomes the phonon displacement operator:
\begin{equation}
v_{{\bf q}\lambda} \to \sqrt{\frac{\hbar}{2 M \omega_{{\bf q}\lambda}}} (a_{{\bf q}\lambda} + a^\dagger_{-{\bf q}\lambda}),
\end{equation}
where $a_{{\bf q}\lambda}$ is an operator that annihilates a phonon mode $\lambda$ with momentum ${\bf q}$.

Since ${\bf w}_{{\bf l}s}(t)$ must be real (or a hermitian operator, in the second quantized version), it follows that ${\bf w}_{{\bf q}s}^*(t)={\bf w}_{-{\bf q}s}(t)$.
This condition implies ${\bf p}_{{\bf q}\lambda s}/\sqrt{\omega_{{\bf q}\lambda}}={\bf p}^*_{-{\bf q}\lambda s}/\sqrt{\omega^*_{-{\bf q}\lambda}}$, which
can be satisfied via ${\bf p}_{{\bf q}\lambda s} = {\bf p}^*_{-{\bf q}\lambda s}$ and $\omega_{{\bf q}\lambda} = \omega^*_{-{\bf q}\lambda}$.

Let us now consider the interaction of lattice vibrations with electromagnetic fields.
In this case, the equations of motions for the lattice vibrations can be obtained from
\begin{equation}
\label{eq:el}
\frac{d}{dt}\left(\frac{\partial L}{\partial \dot{\bf w}_{{\bf l}s}}\right) - \frac{\partial L}{\partial{\bf w}_{{\bf l}s}}=0,
\end{equation}
where $L$ is the full electromechanical Lagrangian,
\begin{equation}
\label{eq:l0}
L=\frac{1}{2}\sum_{{\bf l} s} M_s \dot{{\bf w}}_{{\bf l} s}^2 - \frac{1}{2}\sum_{{\bf l} s j}\sum_{{\bf l}' s' j'} w_{{\bf l} s j} A_{s j, s' j'} ({\bf l}-{\bf l}')w_{{\bf l}'s'j'}
+\int d^3{\bf r}\left[(\grad\times{\bf A})^2 - (\grad\phi+\dot{\bf A})^2\right]+\int d^3{\bf r} ({\bf A}\cdot{\bf j} -\phi \rho),
\end{equation}
where ${\bf A}$ and $\phi$ are the electromagnetic potentials, and  ${\bf j}$ and $\rho$ are the current and charge densities (which include contributions from the lattice).
Combining Eqs.~(\ref{eq:el}) and (\ref{eq:l0}), and assuming for simplicity a point-like ion model, we get
\begin{equation}
\label{eq:phonax}
M_s \ddot{w}_{{\bf l}s j} (t) +\sum_{{\bf l}' s' j'} A_{{\bf l} s j, {\bf l}' s' j'} w_{{\bf l}' s' j'}(t) = e Z_s {\bf E}({\bf R}({\bf l},s)),
\end{equation}
where $e Z_s$ is the charge of ion $s$.
We can write the electric field as 
\begin{equation}
{\bf E}({\bf r}, t) =\sum_{\bf q} \sum_{\bf G} e^{i ({\bf q}+{\bf G})\cdot{\bf r}} {\bf E}_{{\bf q}+{\bf G}}(t),
\end{equation}
where ${\bf q}$ is the momentum inside the first Brillouin zone and ${\bf G}$ are the reciprocal lattice vectors.
It is convenient to separate the electric field into a part ${\bf E}^{(0)}({\bf r})$ that varies slowly on an atomic scale and a part $\delta{\bf E}({\bf r})$ that varies rapidly on an atomic lenghtscale,
\begin{align}
{\bf E}^{(0)}({\bf r}, t) &=\sum_{{\bf q}} e^{i {\bf q}\cdot{\bf r}} {\bf E}_{\bf q}(t)\nonumber\\
\delta{\bf E}({\bf r}, t) &=\sum_{{\bf q}} \sum_{\bf G\neq 0} e^{i ({\bf q}+{\bf G})\cdot{\bf r}} {\bf E}_{{\bf q}+{\bf G}}(t),
\end{align}
In practice, ${\bf E}^{(0)}({\bf r})$ includes the external fields (whose wavelengths exceed greatly the lattice constant in the case of visible or infrared light) as well as unit-cell-averaged internal fields (produced by ionic displacements away from equilibrium).
Unless short-wavelength external probes are used, the rapidly varying part of the electric field originates solely from ionic displacements, and as such it can be absorbed in the left-hand side of Eq.~(\ref{eq:phonax}).   
Thus, we will keep only ${\bf E}^{(0)}$ in the right hand side of the equation of Eq.~(\ref{eq:phonax}).   

With this in mind, Eq. (\ref{eq:phonax}) can be simplified by writing ${\bf w}_{{\bf l} s}$ in terms of the normal modes $v_{{\bf q}\lambda}$ (cf. Eq.~(\ref{eq:wl})), multiplying the two sides of Eq.~(\ref{eq:phonax}) by $\sqrt{N} \exp[-i {\bf q}\cdot{\bf l}] p_{{\bf q}\lambda s j}^*$, summing over ${\bf l},s,j$ and using the orthogonality relations for the polarization vectors.
The outcome reads
\begin{equation}
\label{eq:phonax3}
M (\omega_{{\bf q}\lambda}^2-q_0^2) v_{{\bf q}\lambda}(q_0) = \sqrt{N} {\bf Q}^*_{{\bf q}\lambda}\cdot{\bf E}_{\bf q}(q_0),
\end{equation}
where
\begin{equation}
v_{{\bf q}\lambda}(q_0)=\int dt e^{i q_0 t} v_{{\bf q}\lambda}(t)
\end{equation}
is the Fourier-transform of $v_{{\bf q}\lambda}(t)$ to frequency space,
\begin{equation}
{\bf Q}_{{\bf q}\lambda} = \sum_s e Z_s e^{-i{\bf q}\cdot{\bf t}_s}{\bf p}_{{\bf q}\lambda s} ={\bf Q}^*_{-{\bf q}\lambda}
\end{equation}
is the effective charge vector associated with mode $\lambda$.
In addition, we have made use of 
\begin{equation}
\frac{1}{N} \sum_{\bf l} e^{-i {\bf q}\cdot({\bf l}+{\bf t}_s)} {\bf E}^{(0)}({\bf R}({\bf l},s), t) \simeq \frac{1}{\cal V} \int d^3 {\bf r} e^{-i {\bf q}\cdot{\bf r}} {\bf E}^{(0)} ({\bf r},t)= {\bf E}_{\bf q}(t)
\end{equation}
and $\sum_{\bf l} e^{i ({\bf q}-{\bf q}')\cdot}{\bf l}=N\delta_{{\bf q}{\bf q}'}$ (for both ${\bf q}$ and ${\bf q}'$ inside the first Brillouin zone).
In the long-wavelength limit, ${\bf Q}_{{\bf q}\lambda}$ measures the change in the intracell electric polarization produced by atomic displacements in mode $\lambda$.
It amounts to the projection of Born effective charges onto the mode $\lambda$~\cite{gonze1997}.
The effective charge vanishes for acoustic phonons in the ${\bf q}\to 0$ limit, due to $\sum_s Z_s=0$.
The effective charge also vanishes for non-polar phonon modes, known as infrared inactive.
In contrast, whenever ${\bf Q}_{{\bf q}\lambda}\neq 0$, the associated optical phonon is infrared active.

For later reference, we note that Eq.~(\ref{eq:phonax3}) can be obtained from a phonon-only action
\begin{equation}
\label{eq:s_bare}
S_{\rm ph}^{(0)} = \int\frac{d q_0}{2\pi} \sum_{{\bf q}\lambda}\left[\frac{1}{2} M (q_0^2-\omega_{{\bf q}\lambda}^2) v_{{\bf q}\lambda}(q_0) v_{-{\bf q}\lambda}(-q_0) + \sqrt{N} {\bf Q}_{{\bf q}\lambda}\cdot{\bf E}_{-{\bf q}}(-q_0) v_{{\bf q}\lambda}(q_0)\right]
\end{equation}
via $\delta S_{\rm ph}^{(0)}/\delta v_{{\bf q}\lambda}(q_0) = 0$.
For simplicity, in Eq.~(\ref{eq:s_bare}) we have assumed that the phonon frequencies are real (so that $\omega_{\bf q} = \omega_{-{\bf q}}$).

\section{Electron-phonon interaction \label{app:appB}}

In this section, we detail the formalism of electron-phonon interaction in the context of Weyl semimetals (WSM) and derive Eq. (5) of the main text. 

The lattice displacement from its equilibrium position leads to a
deformation potential $\delta U$ which couples to the electronic
density as
\begin{align}
{\cal H}_{\text{ep}} & =\int_{\mathbf{r}}\Psi^{\dagger}(\mathbf{r})\Psi(\mathbf{r})\delta U(\mathbf{r},t)\label{eq:Hep1}\\
\delta U(\mathbf{r},t) & \simeq\sum_{{\bf l},s}\mathbf{w}_{{\bf l}s}(t)\cdot\frac{\partial U(\mathbf{r}-\mathbf{R}_{{\bf l}s})}{\partial\mathbf{R}_{{\bf l}s}}.\label{eq:DU}
\end{align}
In second quantization, the fermionic
fields $\Psi$ can be expanded into a sum over Bloch functions and creation-annihilation
operators $\left(c_{\mathbf{p}n}^{\dagger},\,c_{\mathbf{p}n}\right)$
\begin{equation}
\Psi({\bf r})=\frac{1}{\sqrt{\mathcal{V}}}\sum_{\text{\textbf{p}} n}\text{e}^{\text{i}\mathbf{p}\cdot\mathbf{r}}u_{\mathbf{p}n}(\mathbf{r})\,c_{\mathbf{p}n},\label{eq:KL1}
\end{equation}
where $u_{\mathbf{p}n}(\mathbf{r})$ is the periodic part of
the Bloch function, $n$ is the band index, and $\mathcal{V}$ the
volume of the sample. Putting (\ref{eq:KL1}) in (\ref{eq:Hep1})
and using the decomposition of atomic displacements in phonon normal
modes (Eq.~(\ref{eq:wl})),
as well as 
$u_{\mathbf{\mathbf{p}}n}(\mathbf{r}+\mathbf{l})=u_{\mathbf{p}n}(\mathbf{r})$,
one gets
\begin{align}
{\cal H}_{\text{ep}} & =\sum_{{\mathbf{p}\mathbf{q}\lambda}}\sum_{n n'} g_{n n'}^{\lambda} ({\bf p}, {\bf q})v_{\mathbf{q}\lambda}(t) c_{\mathbf{\mathbf{p}}n}^{\dagger}c_{\mathbf{p}-\mathbf{q} n'}+h.c.\label{eq:HKL}\\
g_{n n'}^{\lambda} ({\bf p},{\bf q}) & =\frac{1}{\sqrt{N}\mathcal{V_{\text{cell}}}}\int_{\mathbf{r}}\,\text{e}^{-\text{i}\mathbf{q\cdot r}}u_{\mathbf{\mathbf{p}}n}^{*}(\mathbf{r})u_{\mathbf{\mathbf{p}-q} n'}(\mathbf{r})\sum_{s}\mathbf{p}_{\mathbf{q}\lambda s}\cdot\partial_{{\bf t}_s} U_{s}(\mathbf{r}-\mathbf{t}_{s}),\label{eq:gKL}
\end{align}
where ${\bf q}$ is the phonon momentum, ${\bf p}_{{\bf q}\lambda s}$ is a polarization vector and we have used $\sum_{\mathbf{l}}\text{e}^{\text{i}{\bf k}\cdot{\bf l}}=N\delta_{\mathbf{k}0}$ as well as $\mathcal{V_{\text{cell}}}=\mathcal{V}/N$, $N$ being the number
of unit cells in the crystal. 
The product $g_{n n'}^\lambda ({\bf p},{\bf q}) v_{{\bf q}\lambda}$ is intensive (independent of the volume of the crystal).

The low-energy physics in WSM is governed by Weyl nodes, namely isolated points in the
Brillouin zone, around which the electronic dispersion is linear. 
Expanding the Bloch functions around each node
$\tau$ with $\mathbf{p}=\mathbf{k}_{\tau}+\mathbf{k}$, we obtain
\begin{equation}
\Psi(\mathbf{r})=\frac{1}{\sqrt{\mathcal{V}}}\sum_{\tau}\text{e}^{\text{i}\mathbf{k}_\tau\cdot\mathbf{r}}\sum_{|\mathbf{k}|<\Lambda,\sigma} e^{i {\bf k}\cdot{\bf r}} u_{\sigma\tau}(\mathbf{r})\,c_{\mathbf{k}\sigma\tau}+\mathcal{O}(\frac{k}{\Lambda}),\label{eq:Psikp}
\end{equation}
where $\sigma$ is the pseudospin that labels the two degenerate bands at the Weyl nodes, $u_{\sigma\tau}({\bf r})\equiv u_{\mathbf{k}_{\tau}\sigma}({\bf r})$ and
$\Lambda$ is a cutoff under which the linear dispersion of electrons
is valid. From Eqs. (\ref{eq:Hep1}) and (\ref{eq:Psikp}),
\begin{align}
H_{\text{ep}} & \simeq\sum_{\lambda \tau \tau' \sigma \sigma'}\sum'_{{\bf k}{\bf q}}\text{\ensuremath{g_{\sigma\sigma',\tau\tau'}^{\lambda}}(\ensuremath{\mathbf{q}})}v_{\mathbf{q}\lambda}(t)c_{\mathbf{k}\sigma\tau}^{\dagger}c_{\mathbf{k-\mathbf{q+k}_{\tau}-k_{\tau'}}\sigma'\tau'}\label{eq:Hint1}\\
\text{\ensuremath{g_{\sigma\sigma', \tau\tau'}^{\lambda}}(\ensuremath{\mathbf{q}})} & =\frac{1}{\sqrt{N}\mathcal{V_{\text{cell}}}}\int_{\mathbf{r}}\,u_{\sigma\tau}^{*}(\mathbf{r})u_{\sigma'\tau'}(\mathbf{r})\sum_{s}\text{e}^{-\text{i}\mathbf{q}\cdot\mathbf{r}}\mathbf{p}_{\mathbf{q\lambda}s}\cdot\frac{\partial U(\mathbf{r}-\mathbf{t}_s)}{\partial\mathbf{t}_s},\label{eq:gsimp}
\end{align}
where $\sum'$ means that the sum over ${\bf k}$ and ${\bf q}$ is constrained by $|\mathbf{k}|<\Lambda$ and $|\mathbf{\mathbf{k-\mathbf{q+k}_{\tau}-k_{\tau'}}}|<\Lambda$.
Notice that the electron-phonon vertex has been simplified such that its dependence on electronic momenta is contained only in the indices $\tau,\,\tau'$.

From here on, we focus on a single pair of Weyl nodes.
The results obtained below can be generalized to the case of multiple pairs of nodes.
Furthermore, we consider only long wavelength phonons, whose momenta ${\bf q}$ are much smaller than the separation between the nodes. Such phonons
cannot couple distinct nodes and only terms $\tau=\tau'$ remain in Eq.~(\ref{eq:Hint1}).
Then,  
\begin{align}
H_{\text{ep}} & \simeq\sum_{{\bf k}{\bf q}} \sum_{\sigma\sigma'\tau} \left(\sum_\lambda g_{\sigma\sigma',\tau}^{\lambda}({\bf q}) v_{\mathbf{q}\lambda}(t)\right) c_{\mathbf{k}\tau\sigma}^{\dagger}c_{\mathbf{k-q}\tau\sigma'}\label{eq:HKL2},
\end{align}
where $|{\bf k}|<\Lambda$, $|{\bf k}-{\bf q}|<\Lambda$ inside the momenta sums and $g_{\sigma\sigma',\tau}^\lambda\equiv g^\lambda_{\sigma\sigma,\tau\tau}$.
This equation coincides with Eq. (5) of the main text.
We can now decompose $g_{\sigma\sigma',\tau}^{\lambda}$ in the subspace of $\sigma$ and $\tau$, as
\begin{equation}
\label{eq:deco}
g^\lambda_{\sigma\sigma',\tau} = g^\lambda_{0 0}\delta_{\sigma\sigma'} + {\bf g}^\lambda_0 \cdot{\boldsymbol\sigma}_{\sigma\sigma'} + \tau (g^\lambda_{0 z}\delta_{\sigma\sigma'} + {\bf g}^\lambda_z\cdot{\boldsymbol\sigma}_{\sigma\sigma'}),
\end{equation}
where
\begin{align}
\label{eq:gs}
g^\lambda_{00} ({\bf q}) & = \sum_{\sigma\tau} g^\lambda_{\sigma\sigma,\tau}({\bf q})/4\nonumber\\
{\bf g}^\lambda_0 ({\bf q}) &=\sum_{\sigma\sigma'\tau}{\boldsymbol\sigma}_{\sigma\sigma'} g^\lambda_{\sigma\sigma',\tau}({\bf q})/4\nonumber\\
g_{0 z} ({\bf q}) & =\sum_{\sigma\tau}\tau g^\lambda_{\sigma\sigma,\tau}({\bf q})/4\nonumber\\
{\bf g}^\lambda_z({\bf q}) &=\sum_{\sigma\sigma'\tau}\tau {\boldsymbol\sigma}_{\sigma\sigma'} g^\lambda_{\sigma\sigma',\tau}({\bf q})/4.
\end{align}
As mentioned in the main text, the preceding decomposition shows that phonons couple to Weyl fermions as vector and axial-vector fields, namely
as 
\begin{align}
\label{eq:cc}
c^{\mu}({\bf q},t)\equiv &\sum_\lambda\left(g_{00}^{\lambda}({\bf q}),\mathbf{g}_{z}^{\lambda}({\bf q})\right)v_{\mathbf{q}\lambda}(t)\nonumber\\
c_{5}^{\mu}({\bf q},t) \equiv &\sum_\lambda\left(g_{0z}^{\lambda}({\bf q}),\mathbf{g}_{0}^{\lambda}({\bf q})\right)v_{\mathbf{q}\lambda}(t),
\end{align}
respectively.
In the long-wavelength limit, $g_{0z}$ and ${\bf g}_0$ constitute phonon-induced fluctuations of the energy and momentum separation between the two Weyl nodes, respectively. 
In the main text and below, we refer to $g_{0z}$ and ${\bf g}_0$ as {\em axial} electron-phonon couplings.

Let us discuss the symmetry conditions under which the axial electron-phonon coupling $c^\mu_5$ is allowed.
First,  the action of a symmetry operation ${\cal R}$ of the crystal's point group on a Bloch state $|\psi_{{\bf k} n}\rangle$ rotates its wave vector from ${\bf k}$ to ${\cal R}{\bf k}$ (see e.g. Ref. [\onlinecite{powell}]).
Second, the electron density is invariant under operations of the point group. 
Third, spatial inversion and mirror operations reverse the chirality of a Weyl node.
Therefore, if the crystal has either an inversion or a mirror symmetry, we have $\sum_\sigma |u_{\sigma\tau}({\bf r})|^2=\sum_\sigma |u_{\sigma,-\tau}({\bf r})|^2$, which in turn implies $g_{0z}({\bf q})=0$.
In sum, $g_{0z}\neq 0$ is allowed only in enantiomorphic crystals.
An enantiomorphic crystal can be or not be time-reversal symmetric.
For an enantiomorphic crystal with time-reversal symmetry, there are generally at least four Weyl nodes. 
In this case, the pairs of nodes related by time-reversal make {\em additivite} contributions to $g_{0z}$.
On the other hand, ${\bf g}_0\neq 0$ does not require breaking mirror symmetries, because it involves $\sum_\tau (...)$ instead of $\sum_\tau \tau (...)$. 
Yet, because the operation of time reversal flips the spin while preserving the chirality, ${\bf g}_0\neq 0$ is possible only if time-reversal symmetry is broken.
If time-reversal is not broken, $\sum_{\sigma} \sigma |u_{\sigma\tau}|^2=0$ is cancelled between nodes related by time-reversal.

The symmetry requirements for phonon-induced electromagnetic gauge fields are can be similarly deduced. 
In particular, $g_{00}$ is the scalar part of the electron-phonon coupling in the $\sigma$ and $\tau$ space, and is thus generally allowed by symmetry.
When it comes to ${\bf g}_z$, it constitutes a phonon-induced electromagnetic vector potential, and as such it can be nonzero only if the crystal lacks time-reversal symmetry.

In the preceding discussion, we have assumed a single pair of Weyl nodes. 
In the generic situation with $N$ pairs of nodes, the decomposition of Eq.~(\ref{eq:deco}) is no longer complete. The sum over $\tau$ in Eq.~(\ref{eq:gs}) is extended to all nodes and the matrix structure increases to matrices of the fundamental representation of the $SU(N)$ group. 
To see how this arises, 
let us consider the electron-phonon coupling as a $2\times2$ matrix in each node, corresponding to the $\sigma$ Pauli matrices above. 
Assuming time-reversal symmetry, only $\sigma_0$ is allowed. 
The $\tau$ degree of freedom labels the $2N$ nodes. 
Equivalence in the labeling of nodes implies that exchanging Weyl nodes corresponds to unitary transformations in $U(2N)$. 
Assuming only non-degenerate Weyl points, chirality decomposes the set of nodes into two subsets, necessarily well-separated in the Brillouin zone. 
For long-wavelength phonons, the full unitary symmetry is broken down as $U\left(2N\right)\to U\left(N\right)_{R}\times U\left(N\right)_{L}$, where  L and R stand for the chiralities of each set of Weyl fermions.
This can be further decomposed into 
$U\left(1\right)_{V}\times U\left(1\right)_{A}\times SU\left(N\right)_{R}\times SU\left(N\right)_{L}$, where V and A stand for axial and vector Abelian phases.
The Abelian generators may be written as
\begin{equation}
\left(\begin{array}{ccccc}
1\\
 & 1\\
 &  & \ddots\\
 &  &  & 1\\
 &  &  &  & 1
\end{array}\right),\,\left(\begin{array}{cccccc}
1\\
 & \ddots\\
 &  & 1\\
 &  &  & -1\\
 &  &  &  & \ddots\\
 &  &  &  &  & -1
\end{array}\right).
\end{equation} 
The first matrix gives the trivial $U(1)$ phase which, gauged, couples vectorially like electromagnetism. 
This constitutes the generalization of $g^\lambda_{00}$ to many nodes. 
The second matrix gives the generalization of the axial-field-like phonons $g^\lambda_{0z}$ discussed above. 
For these, the consequences of the chiral anomaly we discuss in the two-node case follow without modification.

What about the remaining $SU\left(N\right)_{R}\times SU\left(N\right)_{L}$? These show that, in general, phonons may also couple to electrons in WSM like non-Abelian gauge fields.
In the long-wavelength limit, we can focus on phonons that couple to electrons via the generators of the Cartan-sub-algebra, thus breaking down the full chiral $SU(N)$ symmetries. These remaining generators can always be written in a diagonal basis for each chiral $SU(N)$ algebra.
In this situation, there is the added possibility of having signatures of the \emph{non-Abelian} chiral anomaly in phonons. 
We leave this analysis of the non-Abelian chiral anomaly for future work and hereafter proceed with the simple scenario of a single pair of nodes.




\section{Electronic contribution to the phonon effective action \label{app:appC}}

The objective of this section is to show how Eq.~(\ref{eq:phonax3}) is modified when we incorporate the coupling between phonons and Weyl fermions.
In particular, we are to show additional details for the derivations of Eqs.~(10) and (13) in the main text.

As a preliminary task, we set some notational conventions.
The four-vectors for momentum and position are denoted by $p^\mu=(p_0, v {\bf p})$ and  $x^\mu=(t, {\bf x}/v)$, respectively, where $v$ is the Fermi velocity.
The corresponding four-vectors with a ``lowered index'' are $p_\mu= g_{\mu \nu} p^\nu = (p_0, -v {\bf p})$ and  $x_\mu= g_{\mu\nu} x^\nu = (t, -{\bf x}/v)$,
where $g_{\mu\nu}=g^{\mu\nu}={\rm diag}[1,-1,-1,-1]$ is the metric tensor and Einstein's convention for repeated indices is applied.
The four-vector for the gradient is $\partial_\mu =\partial/\partial x^\mu =  (\partial_t, v \grad)$, or alternatively $\partial^\mu = \partial/\partial x_\mu =  g^{\mu\nu} \partial_\nu = (\partial_t, -v \grad)$.
It follows that $p^\mu=i\partial^\mu$.
Since we have taken $\hbar\equiv 1$, $p$ and $\partial$ have energy units, while $x$ has the units of inverse energy. 
We will restore the $\hbar$ factors n the final results.
The Lorentz-invariant scalar product between momentum and position is given by  $p\cdot x=p_\mu x^\mu =  p_0 t - {\bf p}\cdot{\bf x}$.

The four-vectors we use contain the Fermi velocity $v$ rather than the speed of light $c$.
The motivation for this choice lies in the fact that the phase velocity of massless Dirac fermions {\em in crystals} is $v\ll c$.
Moreover, this choice allows to write the Green's function of the Weyl fermion in a formally identical manner to that in high-energy physics, without having to rescale the momenta:
\begin{equation}
\label{eq:green}
G(p)= \left(\gamma^\mu (p_\mu - a_\mu) -\gamma^\mu \gamma^5 a_{5\mu}\right)^{-1},
\end{equation}
where 
\begin{equation}
a_\mu=A^\mu+c^\mu
\end{equation}
is the vector field (including electromagnetic and phonon contributions) and 
\begin{equation}
a_{5\mu}=b^\mu+c_5^\mu
\end{equation}
 is the axial four-vector (including electronic and phonon contributions, cf Eq.~(\ref{eq:cc})).
In our convention, $A_\mu=(e \phi, -e v {\bf A})$, where $\phi$ and ${\bf A}$ are the electromagnetic potentials.
Accordingly, $a(x)$ and $a_5(x)$ have units of energy.
Incidentally, we note that
\begin{align}
c_\mu (x) &= \sum_{\bf q} e^{i {\bf q}\cdot{\bf r}} c_\mu ({\bf q},t)\nonumber\\
c_{5\mu} (x) &= \sum_{\bf q} e^{i {\bf q}\cdot{\bf r}} c_{5\mu} ({\bf q},t).
\end{align}
Below, we will refer extensively to the Fourier transforms 
\begin{align}
a_\mu (k) &= \int d^4 x e^{i k.x} a_\mu (x)\nonumber\\
a_{5\mu} (k) &= \int d^4 x e^{i k.x} a_{5\mu} (x),
\end{align}
which have units of (energy)$^{-3}$.

As mentioned in the main text, integrating out Weyl fermions results in an additional contribution to the effective action for phonons,
\begin{equation}
\label{eq:ds}
S_{\rm int} = -i\ln\,{\rm det}\left(i\gamma^\mu \partial_\mu - \gamma^\mu a_\mu - \gamma^5 \gamma^\mu a_{5\mu}\right).
\end{equation}
The new equation of motion for phonons is obtained from $\delta S/\delta v_{{\bf q}\lambda}(q_0) = 0$, where $S=S^{(0)}+S_{\rm int}$. 
In order to make analytical progress, we proceed with a perturbative expansion in the electromagnetic fields and the electron-phonon coupling,  $S_{\rm int}\simeq S_2+S_3+...$, where
\begin{align}
\label{eq:delS}
S_2 &= \int_q \Pi_{\mu\nu}(q) \left[a^\mu(q) a^\nu(-q) + a_5^\mu(q) a_5^\nu(-q)\right] \nonumber\\
S_3 &=\int_{k,k'} T_{\alpha\mu\nu}(k,k') a^\mu(k) a^\nu (k') a_5^\alpha(-k-k').
\end{align}
Assuming that the Fermi energy is zero, diagrams with an odd number of $a^\mu$ vanish by charge conjugation symmetry (Furry's theorem).
Corrections due to finite Fermi energy will be ignored, as they are subleading to the main (singular) contribution to be discussed below.

Because we are interested in the effective phonon action in the harmonic approximation, we will keep terms that are up to second order in the lattice displacement in $S_2$ and $S_3$. 
Zeroth order terms in the lattice displacement are purely electromagnetic and do not contribute to the effective phonon action, while third (or higher) order terms in the lattice displacement produce anharmonic phonon effects.

\subsection{The $S_2$ term of the effective action}

The leading term in the perturbative expansion is
\begin{equation}
\label{eq:S2_0}
S_2=\int_q\Pi_{\mu\nu}(q)\left[a^\mu (-q) a^{\nu} (q)+a^\mu_5 (-q) a^\nu_5 (q)\right],
\end{equation}
where
\begin{equation}
\label{eq:Pi}
\Pi_{\mu\nu}(q) =  \int \frac{d^4 p}{(2\pi)^4} {\rm Tr}\left[\gamma_\mu G^{(0)}(p) \gamma_\nu G^{(0)}(p-q) \right]
\end{equation}
is the coupled charge and current response function,
and the fields $a^\mu(q)=A^\mu(q)+c^\mu(q)$ and $a^\nu_5(q)=b^\nu(q)+c_5^\nu(q)$ include phonon contributions (cf. Eq.~(\ref{eq:cc})). 

In Eq.~(\ref{eq:Pi}), $d^4 p = d p_0 d^3 (v {\bf p}) $ has units of (energy)$^4$ and $G^{(0)}(p)=(\gamma^\mu p_\mu)^{-1}$.
Our choice of $p^\mu=(p_0,v {\bf p})$, together with energy and momentum conservation at each vertex of the triangle, leads us to adopt the following four-vectors for the external momenta: $k^\mu=(k_0, v {\bf k})$, $k'^\mu=(k'_0, v {\bf k}')$ and $q^\mu=k^\mu+k'^\mu = (q_0, v {\bf q})$.
Accordingly, $k^2=k_\mu k^\mu\neq 0$ and $k'^2\neq 0$ {\em even for real (on-shell) photons}.
This is the price we need to pay in order to borrow the results from high-energy theory.

Because of the full effective Lorentz invariance of the electronic matter,
as well as gauge invariance, we have~\cite{peskin} $\Pi_{\mu\nu}(q)=\Pi(q^{2}) (q_{\mu}q_{\nu}-g_{\mu\nu}q^{2})$,
were $\Pi(q^{2})$ is the (dimensionless) polarization function that has been already calculated in the literature~\cite{dxiao}. 
Consequently, we arrive at 
\begin{align}
S_{2} & \simeq\sum_{\mathbf{q\lambda}}\int\frac{\text{d}q_{0}}{2\pi}\sqrt{N}v_{\mathbf{q}\lambda}(q_{0})\left[\mathbf{Q}_{\mathbf{q\lambda}}^{(1)}(q_{0})\cdot\left(\mathbf{E}_{-\mathbf{q}}(-q_{0})+{\bf E}^{\rm ph}_{-\bf q}(-q_0)\right)+v\mathbf{Q}_{\mathbf{q\lambda}}^{(2)}\cdot\left(\mathbf{B}_{-\mathbf{q}}(q_{0})+{\bf B}^{\rm ph}_{-\bf q}(-q_0)\right)\right]\nonumber\\
&+\frac{e^2{\cal V}}{\hbar^2 v} \sum_{\bf q}\int\frac{d q_0}{2\pi} \Pi(q^2) \left[|{\boldsymbol \epsilon}_{\bf q}(q_0)|^2 - v^2 |{\boldsymbol\beta}_{\bf q}(q_0)|^2\right] 
\label{eq:S2},
\end{align}
where
\begin{align}
{\bf E}^{\rm ph}_{\bf q}(q_0) &= \frac{i}{e v} \sum_\lambda \left [v {\bf q}\, g_{00}({\bf q})-q_0 {\bf g}_z({\bf q})\right] v_{{\bf q}\lambda}(q_0)\nonumber\\
{\bf B}^{\rm ph}_{\bf q}(q_0) &= \frac{i}{e v} \sum_\lambda \left[{\bf q} \times {\bf g}_z({\bf q})\right]v_{{\bf q}\lambda}(q_0)\nonumber\\
{\boldsymbol\epsilon}_{\bf q}(q_0) &= \frac{i}{e v} \sum_\lambda \left [v {\bf q}\, g_{0z}({\bf q})-q_0 {\bf g}_0({\bf q})\right] v_{{\bf q}\lambda}(q_0)\nonumber\\
{\boldsymbol\beta}_{\bf q}(q_0) &= \frac{i}{e v} \sum_\lambda \left[{\bf q} \times {\bf g}_0({\bf q})\right]v_{{\bf q}\lambda}(q_0)\nonumber\\
\mathbf{Q}_{\mathbf{q\lambda}}^{(1)}(q_{0}) & =-i\frac{e\mathcal{V}}{\hbar v^{2}\sqrt{N}}\Pi(q^{2})\left[q_{0}\mathbf{g}_{z}^{\lambda}({\bf q})-v\mathbf{q}\,g_{00}^{\lambda}({\bf q})\right]\nonumber\\
\mathbf{Q}_{\mathbf{q\lambda}}^{(2)}(q_{0}) & =-i\frac{e\mathcal{V}}{\hbar v^{2}\sqrt{N}}\Pi(q^{2})\, v\mathbf{q}\times\mathbf{g}_{z}^{\lambda}({\bf q}).
\label{eq:s42}
\end{align}
In the derivation of Eq.~(\ref{eq:S2}), we have adopted SI units, restored the $\hbar$ factors, and used $\int_{q}=\int\frac{\text{d}^{4}q}{\left(2\pi\right)}=\frac{v^{3}}{\mathcal{V}}\sum_{\mathbf{q}}\int\frac{\text{d}q_{0}}{2\pi}$.
Also, ${\bf E}^{\rm ph}_{\bf q}(q_0)$ and ${\bf B}^{\rm ph}_{\bf q}(q_0)$ are effective electric and magnetic fields produced by non-axial electron-phonon interactions, whereas 
${\boldsymbol \epsilon}_{\bf q}(q_0)$ and ${\boldsymbol \beta}_{\bf q}(q_0)$ are effective electric and magnetic fields produced by axial electron-phonon interactions.
The second line in Eq.~(\ref{eq:S2}) originates from the second term in Eq.~(\ref{eq:S2_0}). 

Let us discuss the various terms entering Eq.~(\ref{eq:S2}).
First, the terms involving ${\bf E}^{\rm ph}_{\bf q}(q_0)$, ${\bf B}^{\rm ph}_{\bf q}(q_0)$, ${\boldsymbol \epsilon}_{\bf q}(q_0)$ and ${\boldsymbol \beta}_{\bf q}(q_0)$ are second order in the lattice displacement.
Consequently, they constitute phonon self-energy terms that renormalize the phonon spectrum.
We have verified numerically that these self-energy terms produce very small renormalizations of the phonon dispersion.
In contrast, as we shall see later, phonon self-energy terms originating from $S_3$ produce a dramatic impact in the phonon dispersion.

The terms involving ${\bf E}_{\bf q}(q_0)$ and ${\bf B}_{\bf q}(q_0)$ are first order in the lattice displacement, and hence they will appear as driving terms in the phonon equations of motions.
As it turns out, the term involving the magnetic field is negligible compared to the term involving the electric field.
To justify this statement, we begin by recognizing that the magnetic field has a static and a dynamic components.
The static part carries a factor $\delta(q_0)$ with it.
As such, its contribution to Eq.~(\ref{eq:S2}) involves only phonon modes with frequency $q_0=0$. 
These modes can only be acoustic, and moreover their coupling to electrons vanishes.
Hence, the static part of the magnetic field in Eq.~(\ref{eq:S2}) can be neglected.
The dynamical part of the magnetic field is related to the curl of the electric field via Faraday's law: ${\bf B}_{\bf q}(q_0) = {\bf q}\times{\bf E}_{\bf q}(q_0)/q_0$.
The electric field has two parts: an external part coming from the incident electromagnetic waves, and an internal part coming from the lattice vibrations.
The external part is transverse (${\bf E}^{\rm ext}\cdot{\bf q} = 0$), while the internal part is approximately longitudinal (${\bf E}^{\rm int} \times {\bf q} = 0$) because in the regime of interest $c q$ is large compared to the phonon frequency. 
Then, ${\bf B}_{\bf q}(q_0) \simeq {\bf q}\times{\bf E}^{\rm ext}_{\bf q}(q_0)/q_0$.
Now, in SI units, the magnetic field and the electric field in an electromagnetic wave are related by ${\bf B}_{\bf q}(q_0)={\bf E}_{\bf q}(q_0)/c$ and $q_0=c q$.
Hence, $v B = (v/c) E \ll E$, which allows us to neglect the magnetic term in Eq.~(\ref{eq:S2}).
Our numerical calculations show that even the terms involving ${\bf E}_{\bf q}(q_0)$ lead to rather small quantitative changes in the phonon dynamics.

In sum, the contributions from $S_2$ to the phonon dynamics are unimportant with respect to the contributions from $S_3$, which we analyze next.

\subsection{The $S_3$ term of the effective action}

The scattering amplitude appearing in $S_3$ can be written as $T_{\alpha\mu\nu}(k,k')=\Gamma_{\alpha\mu\nu}(k,k')+\Gamma_{\alpha\nu\mu} (k',k)$, where
\begin{equation}
\label{eq:Gamma}
\Gamma_{\alpha\mu\nu}(k, k') = -i \int \frac{d^4 p}{(2\pi)^4} {\rm Tr}\left[\gamma_\mu G^{(0)}(p) \gamma_\nu G^{(0)}(p-k') \gamma_\alpha \gamma^5 G^{(0)}(p+k)\right].
\end{equation}


The amplitude $T_{\alpha\mu\nu}$ can be decomposed into longitudinal and transverse parts~\cite{armallis2009},
\begin{equation}
T_{\alpha\mu\nu}(k,k')=\frac{1}{8\pi^2}\left(T^{(l)}_{\alpha\mu\nu}(k,k') - T^{(t)}_{\alpha\mu\nu} (k,k')\right)
\end{equation}
such that $q^\alpha T^{(l)}_{\alpha\mu\nu}\neq 0$ and $q^\alpha T^{(t)}_{\alpha\mu\nu}=0$. 
Here, we use a slightly different convention from the main text, in order to adapt to the notation of Ref.~[\onlinecite{armallis2009}].
Also, both longitudinal and transverse amplitudes obey the Ward identities for the vector current:
\begin{equation}
\label{eq:ward}
k^\mu T^{(l)}_{\alpha\mu\nu}=k'^\nu T^{(l)}_{\alpha\mu\nu}=k^\mu T^{(t)}_{\alpha\mu\nu}=k'^\nu T^{(t)}_{\alpha\mu\nu}=0.
\end{equation}
These relations ensure the gauge invariance of 
\begin{equation}
S_3 = \int \frac{d^4 k}{(2\pi)^4} \frac{d^4 k'}{(2\pi)^4} T_{\alpha\mu\nu}(k,k') a^\mu(k) a^\nu (k') a_5^\alpha(q),
\end{equation}
which is the contribution from the triangle diagram to the effective action.
By virtue of Eq.~(\ref{eq:ward}), the action $S_3$ remains invariant under a gauge transformation
\begin{equation}
\label{eq:gt}
A^\mu(k)\to A^\mu(k)+i k^\mu \xi(k),
\end{equation}
where $\xi(k)$ is an arbitrary (but differentiable) function of $k$.

The explicit expressions for the longitudinal and trasversal amplitudes are known in the literature~\cite{armallis2009}.
The longitudinal part is fixed by the chiral anomaly:
\begin{equation}
T^{(l)}_{\alpha\mu\nu}(k,k')=w_L(k^2,k'^2,q^2) q_\alpha\epsilon_{\mu\nu\rho\sigma} k^\rho k'^\sigma,
\end{equation}
where $q^2 = q_\mu q^\mu=(q_0)^2 - v^2 |{\bf q}|^2$, $\epsilon_{\mu\nu\rho\sigma}$ is the Levi-Civita tensor and 
\begin{equation}
w_L = -4 i/q^2
\end{equation}
 is the longitudinal form factor responsible for the chiral anomaly.
The transverse amplitude, more complicated, can be conveniently decomposed as
\begin{equation}
T^{(t)}_{\alpha\mu\nu}(k,k')=w_T^{(+)}(k^2, k'^2, q^2) t^{(+)}_{\alpha\mu\nu}(k,k')+w_T^{(-)}(k^2,k'^2,q^2) t^{(-)}_{\alpha\mu\nu}(k,k')+\tilde{w}^{(-)}_T(k^2,k'^2,q^2) \tilde{t}^{(-)}_{\alpha\mu\nu} (k,k'),
\end{equation}
where $t^{(\pm)}_{\alpha\mu\nu}$ and $\tilde{t}^{(-)}_{\alpha\mu\nu}$ are transverse tensors given by
\begin{align}
t^{(+)}_{\alpha\mu\nu} (k, k') &= k_\nu \epsilon_{\mu\alpha \rho\sigma} k^\rho k'^\sigma - k'_\mu\epsilon_{\nu\alpha\rho\sigma} k^\rho k'^\sigma - (k\cdot k') \epsilon_{\mu\nu\alpha\rho} (k-k')^\rho +\frac{k^2+k'^2-q^2}{q^2} q_\alpha \epsilon_{\mu\nu\rho\sigma} k^\rho k'^\sigma\nonumber\\
t^{(-)}_{\alpha\mu\nu} (k,k') &= \left[(k-k')_\alpha - \frac{k^2-k'^2}{q^2} q_\alpha\right] \epsilon_{\mu\nu\rho\sigma} k^\rho k'^\sigma\nonumber\\
\tilde{t}^{(-)}_{\alpha\mu\nu}(k,k') &= k_\nu \epsilon_{\mu\alpha\rho\sigma} k^\rho k'^\sigma + k'_\mu \epsilon_{\nu\alpha\rho\sigma} k^\rho k'^\sigma - (k\cdot k') \epsilon_{\mu\nu\alpha\rho} q^\rho
\end{align}
and  $w_T^{(\pm)}$, $\tilde{w}_T^{(-)}$ are transverse ``form factors'' that depend only on $k^2, k'^2$ and $q^2$,
The analytical expressions for these form factors read \cite{armallis2009}
\begin{align}
\label{eq:armi}
w_T^{(+)} (s_1,s_2,s) &= i\frac{s}{\sigma} +\frac{i}{2\sigma^2}\left[(s_{12}+s_2)(3 s_1^2 +s_1(6 s_{12}+s_2)+ 2 s_{12}^2)\ln\frac{s_1}{s}\nonumber\right.\\ 
&\left.+(s_{12}+s_1)(3 s_2^2+s_2(6 s_{12}+s_1)+2 s_{12}^2)\ln\frac{s_2}{s}\right.\nonumber\\
&\left.+(s_1^2 s_2+ 2 s_{12}^2 s_2 +s_1 (2 s_{12}^2+ 6 s_{12} s_2+s_2^2)) \phi(s_1,s_2)\right]\nonumber\\
w_T^{(-)}(s_1,s_2,s) &= i\frac{s_1-s_2}{\sigma}+\frac{i}{2\sigma^2} \left[-(2(s_2+s_{12}) s_{12}^2 - s_1 s_{12} (3 s_1+4 s_{12}) + s_1 s_2(s_1+s_2+s_{12})) \ln\frac{s_1}{s}\right.\nonumber\\
&\left.+(2(s_1+s_{12}) s_{12}^2 - s_2 s_{12} (3 s_2+ 4 s_{12}) +s_1 s_2 (s_1+s_2+s_{12})) \ln\frac{s_2}{s}\right.\nonumber\\
&\left.+(s_1-s_2)(s_1 s_2 + 2 s_{12}^2) \phi(s_1,s_2)\right]\nonumber\\
\tilde{w}_T^{(-)}(s_1,s_2,s) &=-w_T^{(-)}(s_1,s_2,s),
\end{align}
where $s_1\equiv k^2$, $s_2\equiv k'^2$, $s=q^2$, $s_{12}\equiv k\cdot k'$, $\sigma\equiv s_{12}^2-s_1 s_2$ and
\begin{equation}
\phi(x,y)=\frac{1}{\lambda}\left\{2\left[ Li_2(-\rho x) + Li_2(-\rho y)\right]+\ln\frac{y}{x}\ln\frac{1+\rho y}{1+\rho x}+\ln (\rho x) \ln (\rho y) +\frac{\pi^2}{3}\right\}.
\end{equation}
Here, $x=s_1/s$, $y=s_2/s$, $\rho=2(1-x-y+\lambda)^{-1}$, $\lambda=((1-x-y)^2-4 x y)^{1/2}$ and $Li_2$ is the dilogarithm function.
In addition, in Eq.~(\ref{eq:armi}), we have corrected some typos in Ref. [\onlinecite{armallis2009}].

Let us discuss how the longitudinal and transverse tensors contribute to the effective action.
The longitudinal part can be manipulated as follows:
\begin{equation}
\label{eq:long}
w_L(q)\, q_\alpha a_5^\alpha(q) \epsilon_{\mu\nu\rho\sigma} k^\rho k'^\sigma a^\mu(k) a^\nu(k')  = -\frac{4 i}{q^2} q_\alpha a_5^\alpha(q)\left[ \tilde{{\bf E}}(k)\cdot\tilde{{\bf B}}(k') + \tilde{{\bf E}}(k')\cdot\tilde{{\bf B}}(k)\right],
\end{equation}
where $\tilde{\bf E}(k)={\bf E}(k)+{\bf E}^{\rm ph}(k)$, $\tilde{\bf B}(k)={\bf B}(k)+{\bf B}^{\rm ph}(k)$ and 
\begin{align}
\tilde{\bf E}(k)=\int d^4 x e^{i k\cdot x} \tilde{\bf E}(x)\,\,,\,\,\,
\tilde{\bf B}(k)=\int d^4 x e^{i k\cdot x} \tilde{\bf B}(x),
\end{align}
are the electromagnetic fields associated to $a(k)$. 
Because the way we have defined the electromagnetic gauge fields and the gradient four-vector, $\tilde{\bf E}(k)$ and $\tilde{\bf B}(k)$ have the units of energy$^{-2}$.
In contrast, ${\bf E}_{\bf q}(q_0)$ (${\bf B}_{\bf q}(q_0)$) and ${\bf E}^{\rm ph}_{\bf q}(q_0)$ (${\bf B}^{\rm ph}_{\bf q}(q_0)$) have the SI units of electric and magnetic fields (multiplied by time).
In the derivation of Eq.~(\ref{eq:long}), we have used
\begin{equation}
\epsilon_{\mu\nu\rho\sigma} k^\rho a^\mu(k) k'^\sigma a^\nu(k') = 
-\frac{1}{4}\epsilon_{\mu\nu\rho\sigma} F^{\rho\mu}(k) F^{\sigma\nu}(k')=\tilde{\bf E}(k)\cdot\tilde{\bf B}(k')+\tilde{\bf E}(k')\cdot\tilde{\bf B}(k).
\end{equation}
where
\begin{equation}
F^{\alpha\beta}(k) = \int d^4 x e^{i k\cdot x} F^{\alpha\beta}(x) = \int d^4x e^{i k \cdot x} [\partial^\alpha a^\beta(x)-\partial^\beta a^\alpha(x)] = - i [k^\alpha a^\beta (k) - k^\beta a^\alpha(k)]
\end{equation}
is the electromagnetic tensor in momentum space.

Now, let us look at the transverse parts.
A few manipulations, similar to the ones used in the derivation of Eq.~(\ref{eq:long}), lead to the following relations:
\begin{align}
\label{eq:tra}
t^{(+)}_{\alpha\mu\nu} (k, k') a^\mu(k) a^\nu(k') a_5^\alpha(q) &=\frac{1}{4}\left[\tilde{\bf E}(k)\cdot{\boldsymbol\beta}(q)+{\boldsymbol \epsilon}(q)\cdot\tilde{\bf B}(k)\right] k_\nu a^\nu(k') +\frac{1}{4}\left[\tilde{\bf E}(k')\cdot{\boldsymbol \beta}(q)+{\boldsymbol \epsilon}(q)\cdot\tilde{\bf B}(k')\right] k'_\mu a^\mu(k)\nonumber\\ 
&-  (k\cdot k') \epsilon_{\mu\nu\alpha\rho} (k-k')^\rho a^\mu(k) a^\nu(k') a_5^\alpha(q)\nonumber\\
& + \frac{k^2+k'^2-q^2}{4 q^2} q_\alpha a_5^\alpha(q)\left[ \tilde{\bf E}(k)\cdot\tilde{\bf B}(k') + \tilde{\bf E}(k')\cdot\tilde{\bf B}(k)\right]\nonumber\\
t^{(-)}_{\alpha\mu\nu} (k, k') a^\mu(k) a^\nu(k') a_5^\alpha(q) &=\frac{1}{4} \left[(k-k')_\alpha - \frac{k^2-k'^2}{q^2} q_\alpha\right] a_5^\alpha(q)\left[ \tilde{\bf E}(k)\cdot\tilde{\bf B}(k') + \tilde{\bf E}(k')\cdot\tilde{\bf B}(k)\right]\nonumber\\
\tilde{t}^{(-)}_{\alpha\mu\nu} (k, k') a^\mu(k) a^\nu(k') a_5^\alpha(q) &=\frac{1}{4}\left[\tilde{\bf E}(k)\cdot{\boldsymbol \beta}(q)+{\boldsymbol \epsilon}(q)\cdot\tilde{\bf B}(k)\right] k_\nu a^\nu(k') -\frac{1}{4}\left[\tilde{\bf E}(k')\cdot{\boldsymbol \beta}(q)+{\boldsymbol \epsilon}(q)\cdot\tilde{\bf B}(k')\right] k'_\mu a^\mu(k)\nonumber\\ 
&-  (k\cdot k') \epsilon_{\mu\nu\alpha\rho} q^\rho a^\mu(k) a^\nu(k') a_5^\alpha(q),
\end{align}
where ${\boldsymbol\epsilon}_5(q)=-i v{\bf q} a_{5,0}(q) + i q_0 {\bf a}_5(q)$  and ${\boldsymbol \beta}(q)=i v{\bf q} \times{\bf a}_5(q)$ are the fictitious electric and magnetic fields associated with the axial vector (except for unit conversion factors, they coincide with those of Eq.~(\ref{eq:s42})).
The terms proportional to $(k\cdot k')$ play an essential role in ensuring gauge invariance.
We notice that various terms in Eq.~(\ref{eq:tra}) contain poles at $q^2=0$.
Under certain kinematic configurations~\cite{armallis2009}, these terms can cancel the $1/q^2$ pole from the longitudinal part of the triangle diagram.

We are interested in scenarios where the $1/q^2$ pole of the longitudinal amplitude (which is a hallmark of the chiral anomaly) will not be cancelled by the poles in the transverse amplitude. 
By inspection, we identify two such scenarios: (i) a time-independent and spatially uniform magnetic field ${\bf B}_0$ is applied in the laboratory; (ii) a time-independent and spatially uniform electric field ${\bf E}_0$ is present inside the sample.
Scenario (ii) would involve a charge current and hence would require connecting the sample to a battery. 
Because optical spectroscopy experiments are conducted in absence of transport currents, we will not develop scenario (ii) in this work.
Instead, we will concentrate on scenario (i), which is realizable in optical experiments carried out in static magnetic fields.
 
In the presence of a dominant static and uniform magnetic field, we have 
\begin{equation}
\tilde{\bf B}(k)\simeq{\bf B}_0\int d^4 x e^{i k\cdot x} = (2\pi)^4 {\bf B}_0 \delta^{(4)}(k),
\end{equation}
where $B_0$ has units of energy squared.
The dynamical part of the magnetic field does contribute to the action,  because ${\bf E}(k)\cdot {\bf B}(k')\neq 0$ for electromagnetic waves when $k\neq k'$.

Aside from being small, this part is not interesting because it will not lead to any singular term by virtue of Ref.~[\onlinecite{armallis2009}]. 

Consequently, using $k_\mu \delta^{(4)}(k) \to 0$, we find that the $1/q^2$ terms in Eq.~(\ref{eq:tra}) will vanish.
The remaining terms of the transverse amplitude are non singular, and will lead to negligible effects in the phonon dynamics.
Incidentally, if ${\boldsymbol\beta}(q)=0$ (which is guaranteed if the crystal has time-reversal symmetry), we get 
\begin{align}
t^{(+)}_{\alpha\mu\nu} (k, k') a^\mu(k) a^\nu(k') a_5^\alpha(q) &= 
\frac{(2\pi)^4}{4}\frac{k^2+k'^2-q^2}{q^2} \left[{\bf E}(k)\cdot{\bf B}_0 \delta^{(4)}(k')+{\bf E}(k')\cdot{\bf B}_0 \delta^{(4)}(k)\right] a_5^\alpha(q) = 0\nonumber\\
t^{(-)}_{\alpha\mu\nu} (k, k') a^\mu(k) a^\nu(k') a_5^\alpha(q) &=\frac{(2\pi)^4}{4} \left[(k-k')_\alpha - \frac{k^2-k'^2}{q^2} q_\alpha\right] a_5^\alpha(q)\left[ {\bf E}(k)\cdot{\bf B}_0 \delta^{(4)}(k') + {\bf E}(k')\cdot{\bf B}_0 \delta^{(4)}(k)\right]= 0\nonumber\\
\tilde{t}^{(-)}_{\alpha\mu\nu} (k, k') a^\mu(k) a^\nu(k') a_5^\alpha(q) &= 0.
\end{align}
Therefore, the entire transversal part of the axial action vanishes in this configuration.

In sum, in the presence of ${\bf B}_0\neq 0$, 
the contribution to $S_3$ is dominated from the chiral anomaly pole in the longitudinal part of the triangle diagram.
Thus, we may approximate
\begin{equation}
S_3\simeq-\frac{i}{\pi^2} \int \frac{d^4 q}{(2\pi)^4} \frac{q_\alpha}{q^2} a_5^\alpha(q) \tilde{\bf E}(q)\cdot{\bf B}_0,
\end{equation}
where $a_5$ includes a static (band structure) part and a dynamic (phonon) part.
The contribution from the dynamic part to $S_3$ will be denoted as $S_3^{(a)}$, while the contribution from the static part will be denoted by $S_3^{(b)}$, such that 
$S_3=S_3^{(a)}+S_3^{(b)}$. 
On one hand,
\begin{align}
\label{eq:S3a}
S_3^{(a)} & =-\frac{i}{\pi^{2}\hbar^{2}}\int_{q}\tilde{\bf E}(q)\cdot\mathbf{B}_{0}\frac{q_\alpha c_5^\alpha(-q)}{q^{2}}\nonumber \\
 & =\sqrt{N}\sum_{\mathbf{q}\lambda}\int\frac{\text{d}q_{0}}{2\pi}\,\left(\mathbf{E}_{\mathbf{q}}(q_{0})+{\bf E}^{\rm ph}_{\bf q}(q_0)\right)\cdot\delta\mathbf{Q}_{-\mathbf{q},\lambda}(-q_{0})\,v_{-\mathbf{q} \lambda}(-q_{0})\\
\delta\mathbf{Q}_{-\mathbf{q}\lambda}(-q_{0}) & =-i\frac{e^{2}\mathcal{V}}{\hbar^{2}\pi^{2}\sqrt{N}}\frac{q_0 g^\lambda_{0z}({\bf q})-v {\bf q}\cdot{\bf g}^\lambda_0({\bf q})}{q^{2}}\mathbf{B}_{0}\label{eq:dQ}
\end{align}
On the other hand, 
\begin{align}
S^{(b)}_{3} & \simeq\sum_{\mathbf{q\lambda}}\int\frac{\text{d}q_{0}}{2\pi}\sqrt{N}v_{\mathbf{q}\lambda}(q_{0})\left[\mathbf{Q}_{\mathbf{q\lambda}}^{'(1)}(q_{0})\cdot\left(\mathbf{E}_{-\mathbf{q}}(-q_{0})+{\bf E}^{\rm ph}_{\bf q}(q_0)\right)+v\mathbf{Q}_{\mathbf{q\lambda}}^{'(2)}\cdot\left(\mathbf{B}_{-\mathbf{q}}(q_{0})+{\bf B}^{\rm ph}_{\bf q}(q_0)\right)\right]\label{eq:S3b}\\
\mathbf{Q}_{\mathbf{q\lambda}}^{'(1)}(q_{0}) & =-\frac{e\mathcal{V}}{\pi^{2}\hbar^{2}v^{2}\sqrt{N}}b^{\mu}\frac{\partial}{\partial q_{\mu}}\left(v\mathbf{q}\times\mathbf{g}_{z}^{\lambda}({\bf q})\right)\\
\mathbf{Q}_{\mathbf{q\lambda}}^{'(2)}(q_{0}) & =-\frac{e\mathcal{V}}{\pi^{2}\hbar^{2}v^{2}\sqrt{N}}b^{\mu}\frac{\partial}{\partial q_{\mu}}\left(q_{0}\mathbf{g}_{z}^{\lambda}({\bf q})-v\mathbf{q}g_{00}^{\lambda}({\bf q})\right).
\end{align}
The contribution from Eq.~(\ref{eq:S3a}) to the phonon dynamics is remarkable because of the $1/q^2$ anomaly pole contained in Eq.~(\ref{eq:dQ}).
In comparison, the contribution from Eq.~(\ref{eq:S3b}) is minor and its omission does not incur in any qualitative errors.
In Eq.~(\ref{eq:S3a}), the term involving ${\bf E}^{\rm ph}_{\bf q}(q_0)$ is of second order in the phonon displacement and leads to an anomalous renormalization of the phonon dispersion.
The term containing ${\bf E}^{\rm ph}_{\bf q}(q_0)$ is of first order in the phonon displacement.
As we shall show in the next section, this term can lead to both an anomalous optical absorption as well as an anomalous renormalization of the phonon dispersion. 

\section{Lattice dynamics in the presence of coupling to  Weyl fermions\label{app:appD}}

The equation of motion for phonons is obtained from $\delta S/\delta v_{{\bf q}\lambda}(q_0) = 0$, where $S=S^{(0)}+S_{\rm int}$. 
Using Eqs.~(\ref{eq:S2}), (\ref{eq:S3a}) and (\ref{eq:S3b}), we obtain 
\begin{equation}
M\left(\omega_{\mathbf{q\lambda}}^{2}-q_0^2\right)v_{\mathbf{q\lambda}}(q_{0})\simeq\sqrt{N}\mathbf{Q}_{\mathbf{q\lambda}}(q_{0})^{*}\cdot\left(\mathbf{E}_{\bf q}(q_0)+{\bf E}^{\rm ph}_{\bf q}(q_0)\right),\label{eq:edmph1}
\end{equation}
where $\mathbf{Q}_{\mathbf{q\lambda}}(q_{0})\simeq \mathbf{Q}_{\mathbf{q\lambda}}^{(0)}(q_{0})+\delta\mathbf{Q}_{\mathbf{q}\lambda}(q_{0})$ and we have neglected all non-singular contributions whose impact in the phonon dynamics is demonstrably minor.

The ${\bf E}^{\rm ph}_{\bf q}(q_0)$ term appearing in Eq.~(\ref{eq:edmph1}) is linear in the phonon displacement and can thus be reabsorbed on the left hand side. 
On the other hand, the electric field ${\bf E}_{\bf q}(q_0)$ contains an internal fluctuating part due to lattice vibrations and an external part coming from photons incident in the sample.
The external field is transverse (${\bf q}\cdot {\bf E}_{\bf q}^{\text{ext}}=0$),  whereas the internal field is approximately longitudinal (${\bf q}\times {\bf E}_{\bf q}^{\text{int}}=0$).
We thus set ourselves in a regime where Coulomb retardation effects are unimportant ($c |{\bf q}| \gg \omega$).
In this regime, the usual polariton effects are absent.
However, as we shall see, there will be novel polariton effects that will emerge due to the chiral anomaly.



In order to obtain an expression for the internal field, we use Maxwell's equations. 
From Gauss' law, we have
\begin{equation}
\label{eq:gauss}
\hat{\mathbf{q}}\cdot\mathbf{E}_{\bf q}(q_0)=\hat{\mathbf{q}}\cdot\left(\mathbf{E}^{\text{int}}_{\bf q}(q_0)+\mathbf{E}^{\text{ext}}_{\bf q}(q_0)\right)=\hat{\bf q}\cdot {\bf E}_{\bf q}^{\rm int}(q_0)=  -\frac{1}{\epsilon_0}\hat{\mathbf{q}}\cdot\mathbf{P}_{\bf q}(q_0).
\end{equation}
Here, the polarization ${\bf P}$ is defined by ${\bf j}=\partial {\bf P}/\partial t$, where ${\bf j}$ is the total current including ionic displacements as well itinerant carriers.
Next, we use the constitutive relation ${\bf P}_{\bf q}(q_0)=\epsilon_0 \chi({\bf q}, q_0) {\bf E}_{\bf q}(q_0)$, where $\chi$ is the total (electronic and lattice) dimensionless susceptibility.
For simplicity, we assume an isotropic medium so that $\chi$ is a scalar.
Through the definition of the mode-effective phonon charge as the change in the intracell polarization due to lattice vibrations, we may write
\begin{equation}
\label{eq:P}
{\bf P}_{\bf q}(q_0)=\frac{1}{{\cal V}_{\rm cell} \sqrt{N}}\sum_\lambda {\bf Q}_{{\bf q}\lambda}(q_0) v_{{\bf q}\lambda}(q_0) + \epsilon_0 \chi_e({\bf q},q_0) {\bf E}_{\bf q}(q_0),
\end{equation}
where $\chi_e$ is the electronic susceptibility, related to the electronic dielectric function $\epsilon_e({\bf q},q_0)$ via $\chi_e =\epsilon_e/\epsilon_0-1$.  
To simplify the discussion, we have considered a crystal with time-reversal symmetry (${\bf g}_0={\bf g}_z={\bf b}=0$).
Hence, Eq.~(\ref{eq:P}) does not have a magnetoelectric coupling in the electronic sector.
The prefactor $1/\sqrt{N}$ in the first term of Eq.~(\ref{eq:P}) follows from the conventions we have adopted: $v_{{\bf q}\lambda}$ scales like $\sqrt{N}$ with system size, while ${\bf P}_{\bf q}$, ${\bf E}_{\bf q}$ and ${\bf Q}_{{\bf q}\lambda}$ are intensive quantities.
Also, we have included a factor of the unit cell volume as per the usual definition of the Born effective charge~\cite{gonze1997}.
Combining Eqs.~(\ref{eq:P}) and (\ref{eq:gauss}), we obtain
\begin{equation}
E^{\text{int}}_{\bf q}(q_0)=  -\frac{1}{{\cal V}_{\rm cell} \sqrt{N} \epsilon_0}\sum_{\lambda}\left(\hat{\mathbf{q}}\cdot\mathbf{Q}_{\mathbf{q}\lambda}\right)v_{\mathbf{q}\lambda}+(1-\epsilon_e/\epsilon_0)E^{\text{int}}_{\bf q}(q_0)
\Rightarrow
\mathbf{E}^{\text{int}}_{\bf q}(q_0)=  -\hat{\mathbf{q}}\frac{1}{\epsilon_e {\cal V}_{\rm cell} \sqrt{N}}\sum_{\lambda}\left(\hat{\mathbf{q}}\cdot\mathbf{Q}_{\mathbf{q}\lambda}\right)v_{\mathbf{q}\lambda}.
\end{equation}
Including this in (\ref{eq:edmph1}), we find the phonon dispersion is modified by a self-energy term,
\begin{align}
M\sum_{\lambda'}\mathcal{D}_{\lambda\lambda'}({\bf q},q_0)v_{\mathbf{q\lambda}'}(q_0) & =\sqrt{N}\mathbf{Q}_{\mathbf{q}\lambda}^{*}(q_0)\cdot\mathbf{E}^{\text{ext}}_{\bf q}(q_0)\label{eq:edmfin}\\
\text{with }\mathcal{D}_{\lambda\lambda'}({\bf q},q_0) & =\left[\left(\omega_{\mathbf{q}\lambda}^{2}-q_0^2\right)\delta_{\lambda\lambda'}-\frac{e {\cal V}}{M \hbar^2 \pi^2} \frac{g_{0z}^\lambda ({\bf q})\, g_{00}^{\lambda'}({\bf q})\,q_0\, {\bf q}\cdot{\bf B}_0}{q^2}+\frac{\hat{\mathbf{q}}\cdot\mathbf{Q}_{\mathbf{q}\lambda}^{*}(q_0)\, \hat{\mathbf{q}}\cdot\mathbf{Q}_{\mathbf{q}\lambda'}(q_0)}{M\epsilon_e {\cal V}_{\rm cell}}\right].
\end{align}
The condition ${\rm det}\left[{\cal D}_{\lambda\lambda}({\bf q},q_0)\right]=0$ gives the dispersion relation for phonons ($q_0$ vs. ${\bf q}$).
Hereafter, we will be interested in long wavelength optical phonons, which are the ones participating in infrared reflectivity and Raman scattering. 
Although the mode-effective charges vanish for acoustic phonons as ${\bf q}\to 0$, the anomaly at $q^2=0$ ($v^2|{\bf q}|^2 = q_0^2$) may also have an impact in the dispersion of acoustic phonons at finite momenta.
Herein, we will restrict our attention to optical phonons.


Ordinary longitudinal modes are characterized by ${\bf Q}_{\bf q}^{(0)} || \hat{\bf q}$ when ${\bf q}$ points along a high symmetry direction of the crystal.
For transverse modes, ${\bf Q}_{\bf q}^{(0)} \perp \hat{\bf q}$.
This is no longer the case in the presence of a magnetic field, which modifies the direction of the mode-effective phonon charge.
When ${\bf B}_0\neq 0$, a phonon mode that was originally transverse can acquire a total charge that is no longer orthogonal to ${\bf q}$.
This leads to a magnetic-field-induced hybridization of longitudinal and transverse optical phonons.
To highlight the importance of the magnetic field in the phonon spectrum, let us concentrate on a single phonon mode that couples axially to electrons ($g_{0z}\neq 0$). 
We will assume that the frequency of this mode is well-separated from those of the other modes. 
Then, ${\cal D}$ reduces to a scalar and the solutions of ${\cal D}_{\lambda\lambda}=0$ are depicted in Figure 1 of the main text.

In that figure, the $1/q^2$ pole coming from the chiral anomaly leads to the appearance of a mode with $q_0\simeq v |{\bf q}|$, which follows the dispersion of Weyl particle-hole excitations.
This quasilinear mode, which has a composite electronic-phononic-electromagnetic character, hybridizes with an optical phonon when $v|{\bf q}|$ becomes comparable to the frequency of the bare optical phonon.
This hybridization contains polariton-like features whose strength scales as a power law of $\hat{\bf q}\cdot{\bf B}_0$, and can thus be controlled by the magnitude and direction of the magnetic field. In particular,  if we neglect the term linear in $\bf{B}_0$ in Eq. (\ref{eq:eomnum}), the gap between the optical phonon and the quasilinear mode at $v|{\bf q}|=\omega_0$ goes like $|{\hat{\bf q}\cdot{\bf B}_0}|^\frac{2}{3}$.We have verified this power law analytically and numerically.
For completeness, we have considered more complex situations with multiple phonon modes and intermode coupling; these cases too display hybridizations between the Weyl fermions and the optical phonons (not shown).

In the derivation of Figure 1 of the main text, we have carried out the following estimates of all parameters entering Eq.~(\ref{eq:edmfin}). 
For $g_{00}$, we have used $\int_{\bf r}\sum_{\tau\sigma}\left|u_{\tau\sigma}(\mathbf{r)}\right|^{2} \frac{\partial U(\mathbf{r}-\mathbf{t}_{s})}{\partial\mathbf{t}_{s}}\sim\frac{I_{0}}{a_{B}} {\cal V}_{\rm cell}$, where $I_0$ is a Rydberg and $a_B$ is the Bohr radius. 
Then, $\sqrt{N}g_{00}\sim\frac{I_{0}}{a_{B}}\sim 5\times 10^{-8}\text{kg\,m}^{-1}\text{s}^{-2}$.
Since $g_{0z}$ implies a difference between amplitudes of wave functions of different chiralities instead of a sum, we have chosen $g_{0z}\sim 0.1g_{00}$.
The term in Eq.~(\ref{eq:edmfin}) that is linear in $\delta\mathbf{Q}$ (without $\mathbf{g}_{0}$) then reads 
\[
-\frac{\sqrt{N}g_{00}\sqrt{N}g_{0z}}{M}\frac{eV_{\rm cell}}{\pi^{2}\hbar^{2}}\frac{1}{v}\hat{\mathbf{q}}\cdot\mathbf{B}_{0}\frac{q_{0}\,v\left|\mathbf{q}\right|}{q_{0}^{2}-v^{2}\mathbf{q}^{2}}
\]
with $M\sim10^{-25}\text{kg}$, $v\sim 10^5 {\rm m/s}$,  $V_{\rm cell}\sim125\,\mathring{\text{A}}^{3}\sim10^{-28}\text{m}^{3}$, such that
\begin{align}
 & \frac{\sqrt{N}g_{00}\sqrt{N}g_{0z}}{M}\frac{eV_{\rm cell}}{\pi^{2}\hbar^{2}}\frac{1}{v}\sim10^{24}\text{s}^{-2}\text{T}^{-1}\to 1\,\text{meV}^{2}\text{T}^{-1} \label{eq:OOMlin}
\end{align}
The term in Eq.~(\ref{eq:edmfin}) quadratic in $\delta{\bf Q}$ reads
\[
\frac{e^{4}V_{\rm cell}}{M\epsilon_{e}\pi^{4}\hbar^{4}}\left|\sqrt{N}g_{0z}\right|^{2}\left(\mathbf{\hat{q}}\cdot\mathbf{B}_{0}\right)^{2}\left(\frac{q_{0}}{q_{0}^{2}-v^{2}\mathbf{q}^{2}}\right)^{2}
\]
with $\epsilon_{e}\sim10\epsilon_{0}$, which leads to
\begin{align}
 & \frac{e^{4}V_{\rm cell}}{M\epsilon_{e}\pi^{4}\hbar^{4}}\left|\sqrt{N}g_{0z}\right|^{2}\sim10^{48}\text{s}^{-4}\text{T}^{-2}\to 1\,\text{meV}^{4}\text{T}^{-2}.\label{eq:OOMquad}
\end{align}

One is then led to the equation 
\begin{equation}
q_{0}^{2}-\omega_{0}^{2}+K_{lin}\frac{q_{0}\,v\left|\mathbf{q}\right|}{q_{0}^{2}-v^{2}\left|\mathbf{q}\right|^{2}}-K_{quad}\frac{q_{0}^{2}}{\left(q_{0}^{2}-v^{2}\left|\mathbf{q}\right|^{2}\right)^{2}}=0\label{eq:eomnum},
\end{equation}
where $K_{lin}=1 {\rm meV}^2 {\rm T}^{-1} (\hat{\bf q}\cdot{\bf B}_0)$ and $K_{quad}=1 {\rm meV}^4 {\rm T}^{-2} (\hat{\bf q}\cdot{\bf B}_0)^2$.
Figure \ref{fig:manyT} displays the real and imaginary parts of the dispersion for $\mathbf{\hat{q}}\cdot\mathbf{B}_{0}=-0.1\text{T},\;-1\text{T}$
and $-5\text{T}$, respectively, and $\hbar\omega_{0}=10\text{meV}$.
We notice the the imaginary part of the dispersion has weak dependence
on the bare phonon frequency $\omega_{0}$ and rather scales with
the other terms given in equation (\ref{eq:eomnum}). For a small
magnetic field, the linear term $K_{lin}$ is dominant and the imaginary
part is of the order of $\sqrt{K_{lin}}$. For $\mathbf{\hat{q}}\cdot\mathbf{B}_{0}\sim-1\text{T}$,
it scales as $K_{quad}^{1/4}\sim\sqrt{K_{lin}}$. For high magnetic
fields, the quadratic term is dominant and the imaginary part of the dispersion is of the
same order as $K_{quad}^{1/4}$.
A numerical fit shows that the anticrossing between the optical phonon and the pseudoscalar mode  (i.e. the gap between the yellow and red curves at $|{\bf q}|\simeq \omega_0/v$ in Fig.~\ref{fig:manyT}) scales as $|\hat{\bf q}\cdot{\bf B}_0|^{1/2}$.

\begin{figure}[h!]
\mbox{\subfigure{ \includegraphics[width=0.5\columnwidth]{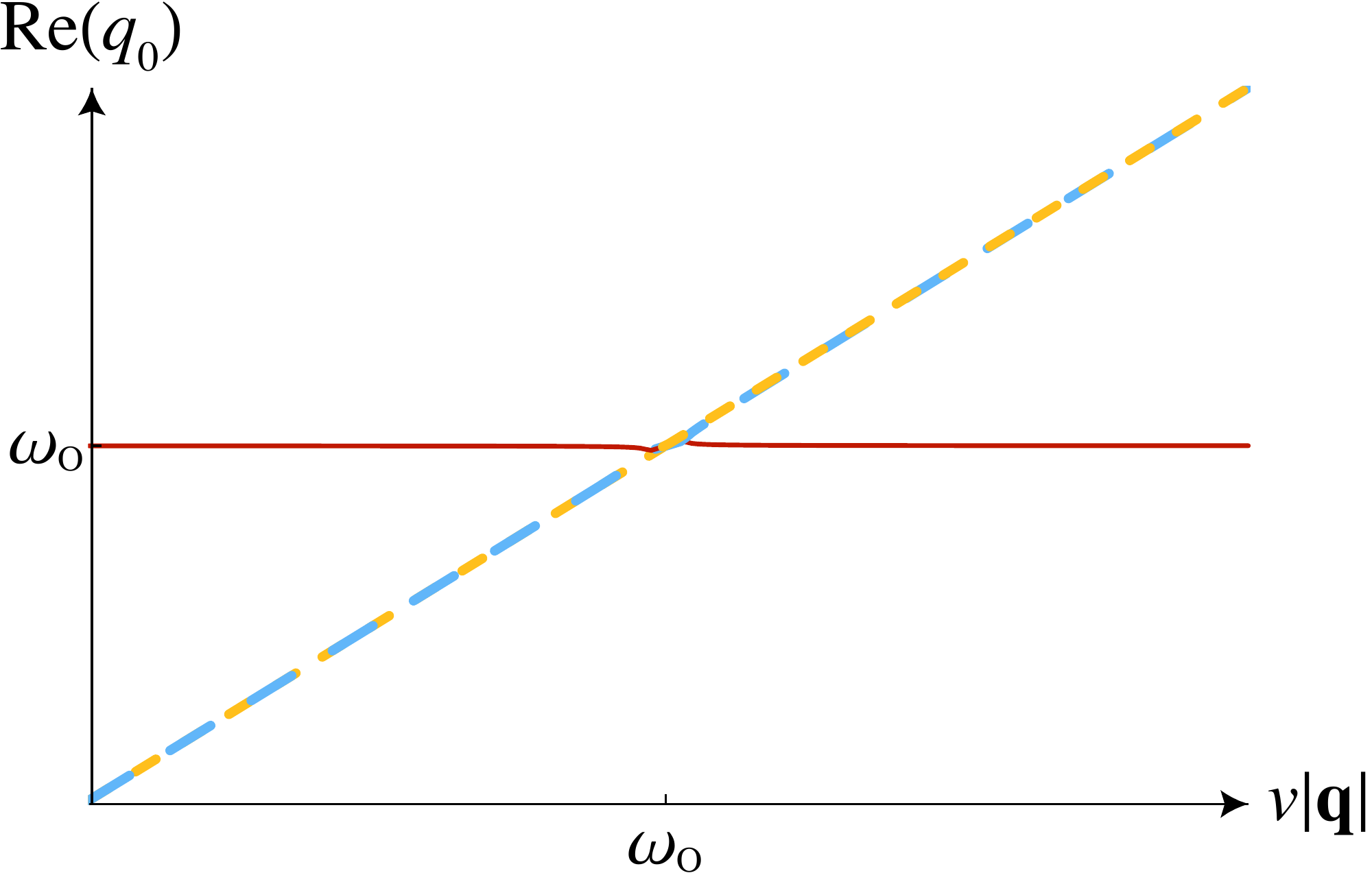}\llap{ \put(-120,0){\raisebox{0.7cm}{\fbox{\includegraphics[width=0.165\columnwidth]{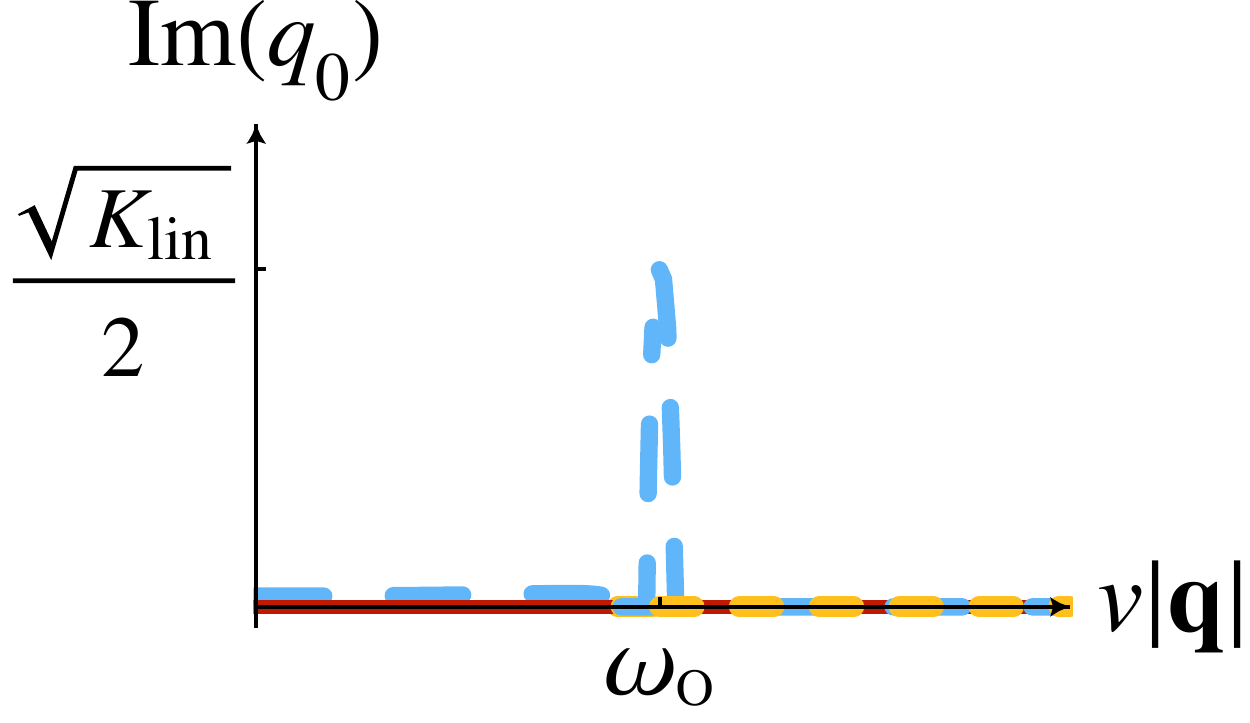}}}}}}\llap{\raisebox{4cm}{\shiftleft{7cm}{$\hat{\bf q}\cdot{\bf B}_0=0.1\text{T}$}}}
\subfigure{ \includegraphics[width=0.5\columnwidth]{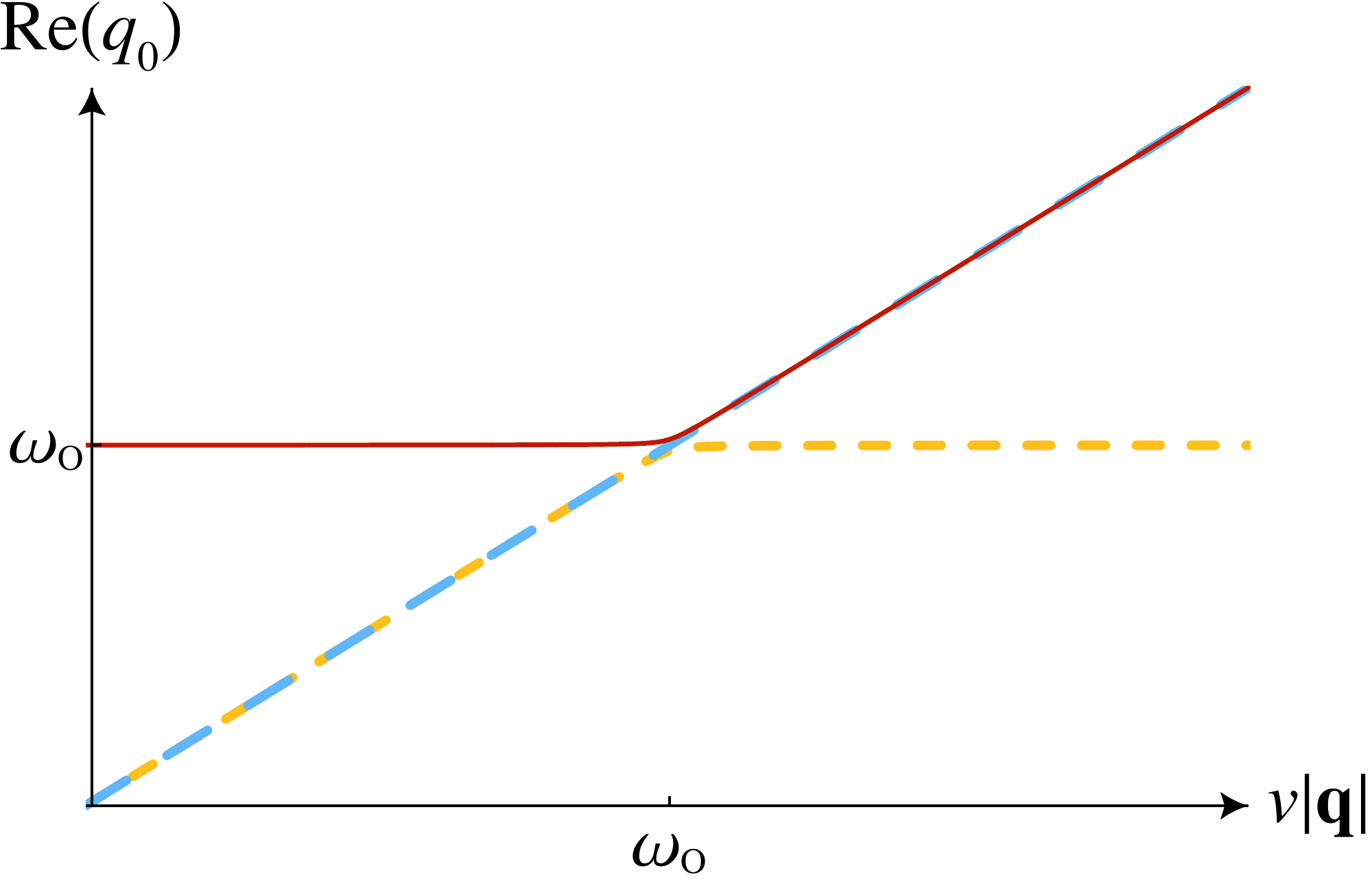}\llap{ \put(-120,0){\raisebox{0.7cm}{\fbox{\includegraphics[clip=true, trim= 0 4.5cm 0 0, width=0.165\columnwidth]{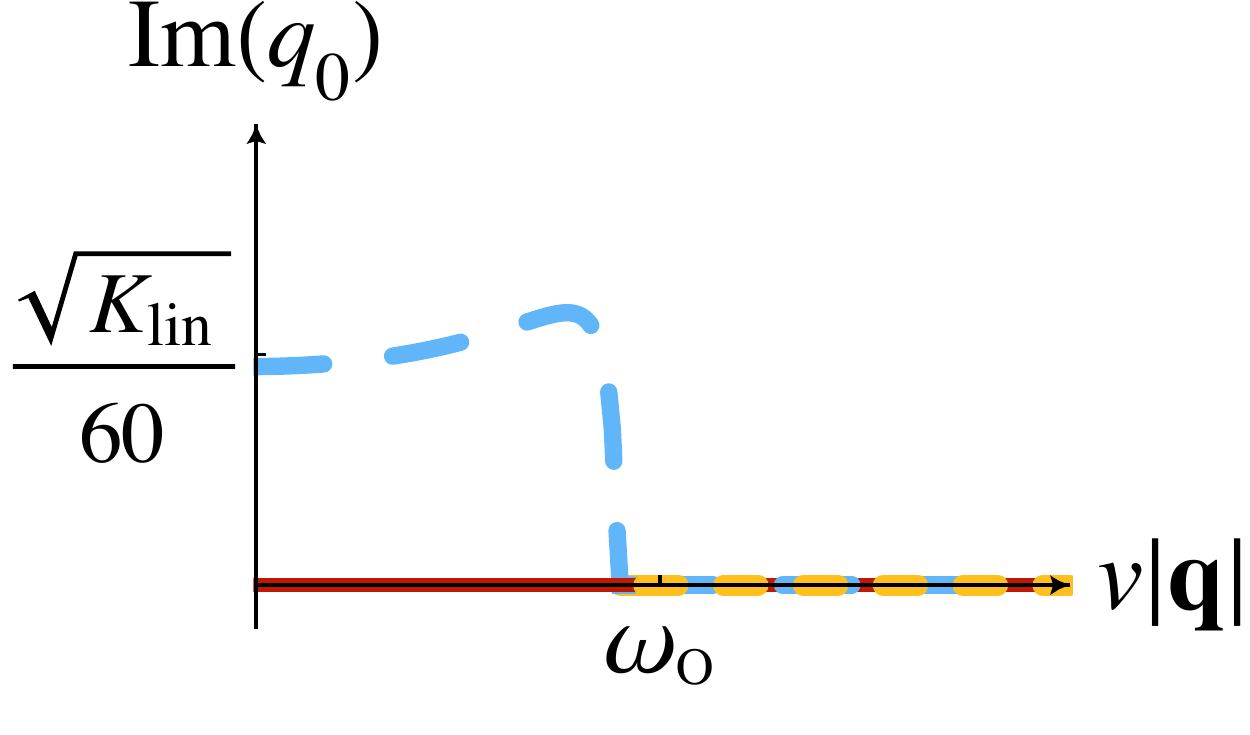}}}}}}\llap{\raisebox{4cm}{\shiftleft{7cm}{$\hat{\bf q}\cdot{\bf B}_0=-0.1\text{T}$}}}
}

\mbox{\subfigure{ \includegraphics[width=0.5\columnwidth]{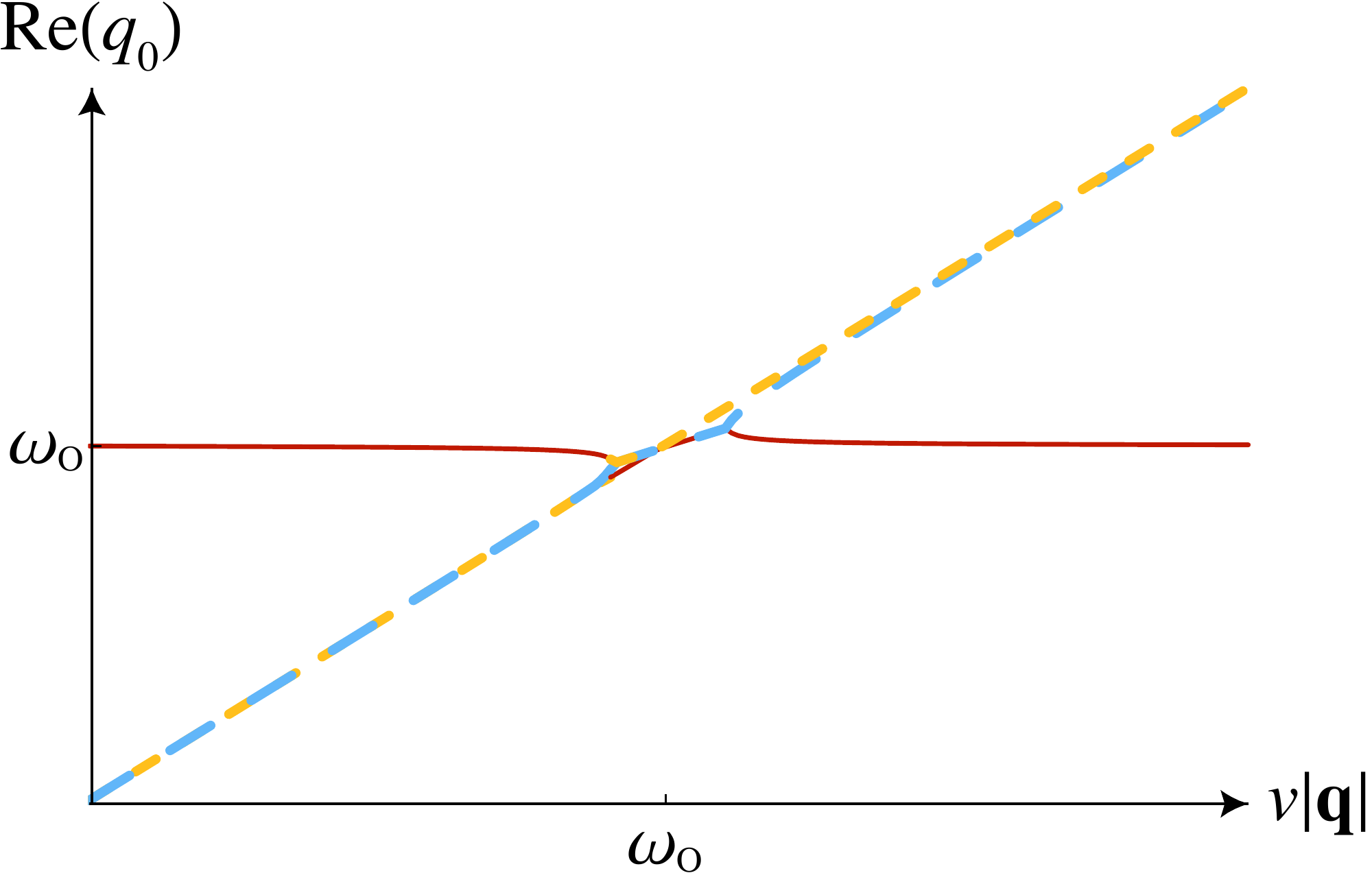}\llap{ \put(-120,0){\raisebox{0.7cm}{\fbox{\includegraphics[width=0.165\columnwidth]{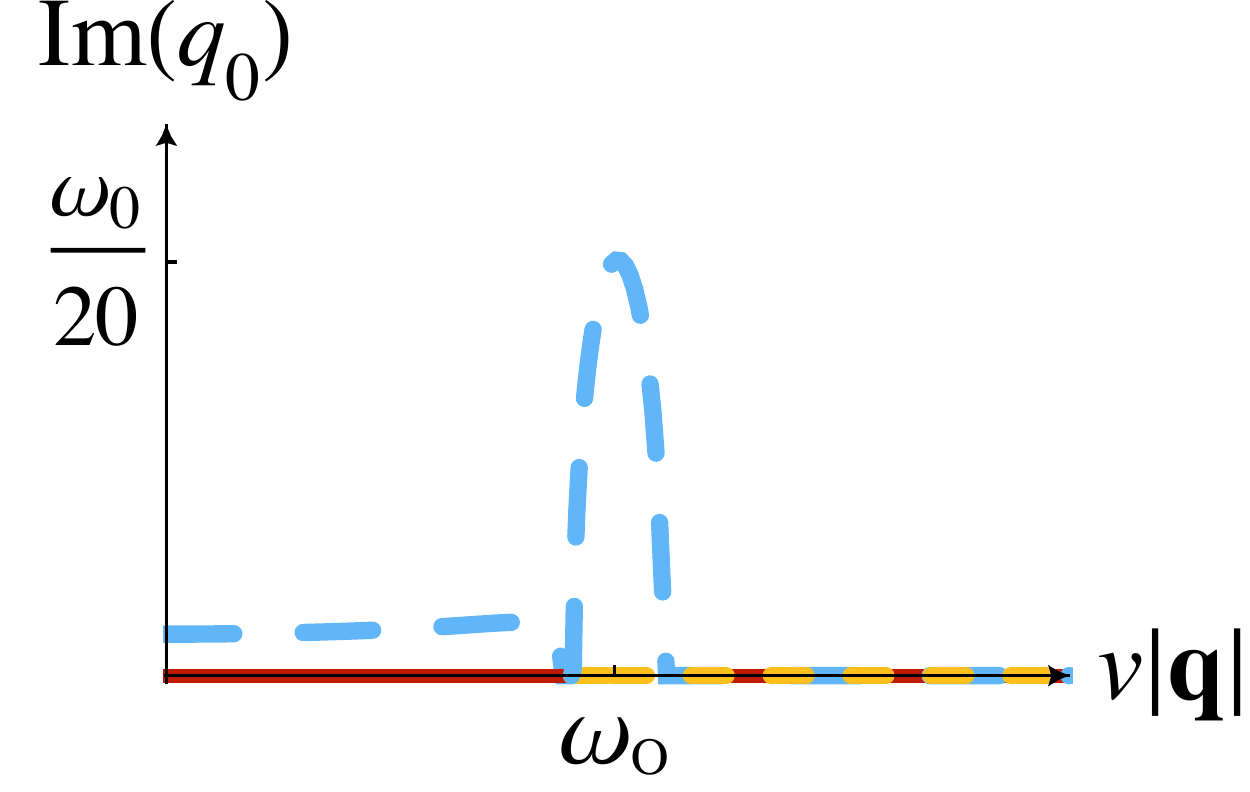}}}}}}\llap{\raisebox{4cm}{\shiftleft{7cm}{$\hat{\bf q}\cdot{\bf B}_0=1\text{T}$}}}
\subfigure{ \includegraphics[width=0.5\columnwidth]{figpho_RE.pdf}\llap{ \put(-120,0){\raisebox{0.7cm}{\fbox{\includegraphics[width=0.165\columnwidth]{figpho_IM.pdf}}}}}}\llap{\raisebox{4cm}{\shiftleft{7cm}{$\hat{\bf q}\cdot{\bf B}_0=-1\text{T}$}}}

}

\mbox{\subfigure{ \includegraphics[width=0.5\columnwidth]{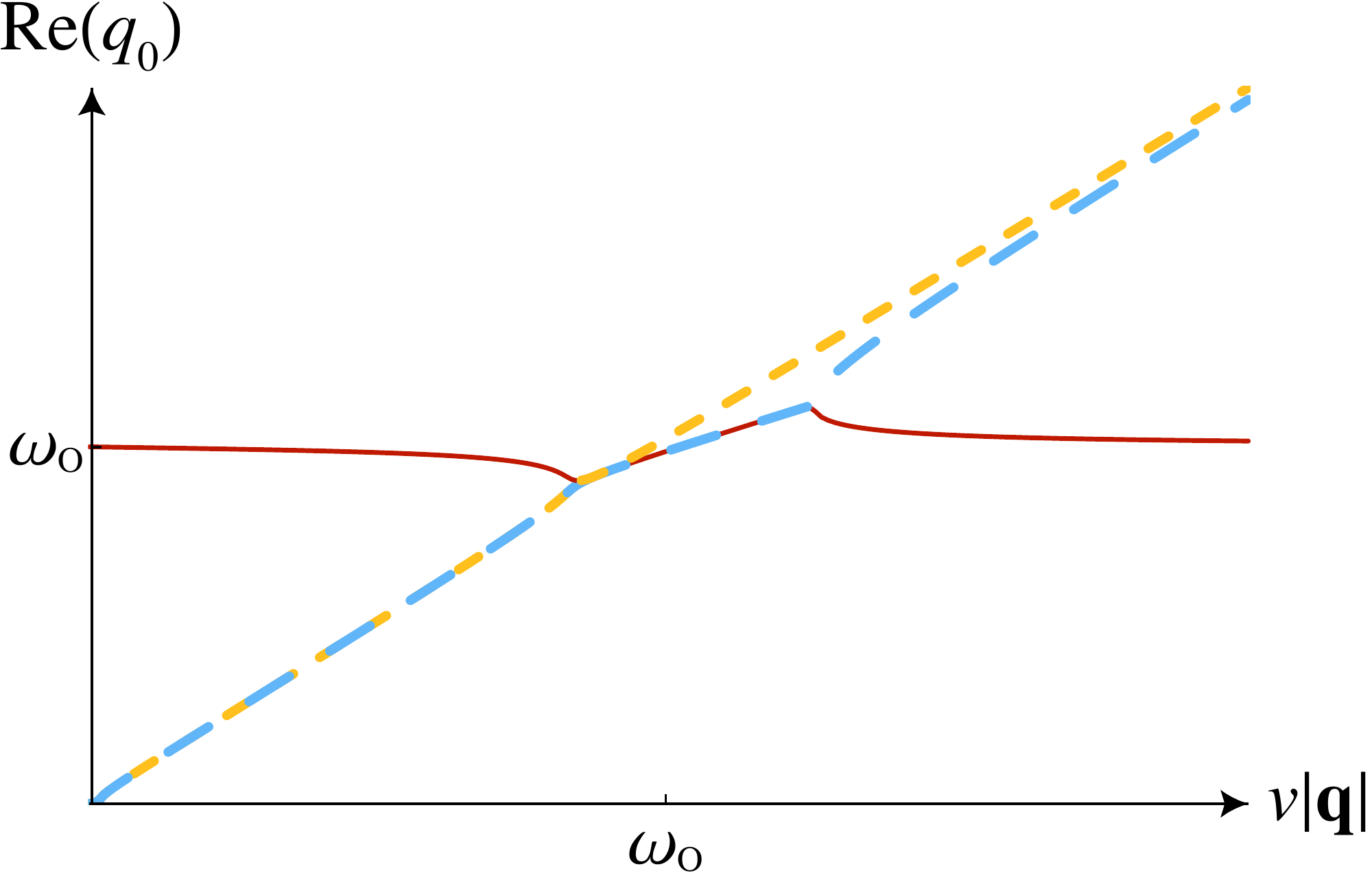}\llap{ \put(-120,0){\raisebox{0.7cm}{\fbox{\includegraphics[width=0.165\columnwidth]{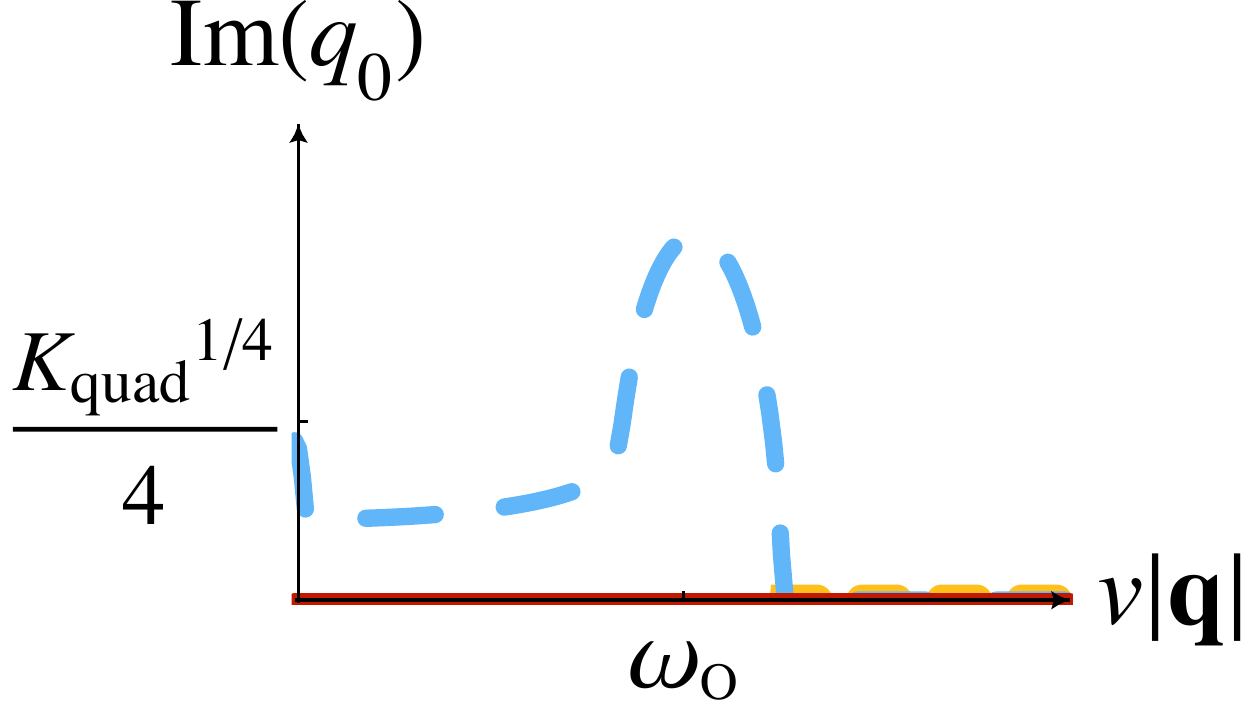}}}}}}\llap{\raisebox{4cm}{\shiftleft{7cm}{$\hat{\bf q}\cdot{\bf B}_0=5\text{T}$}}}
\subfigure{ \includegraphics[width=0.5\columnwidth]{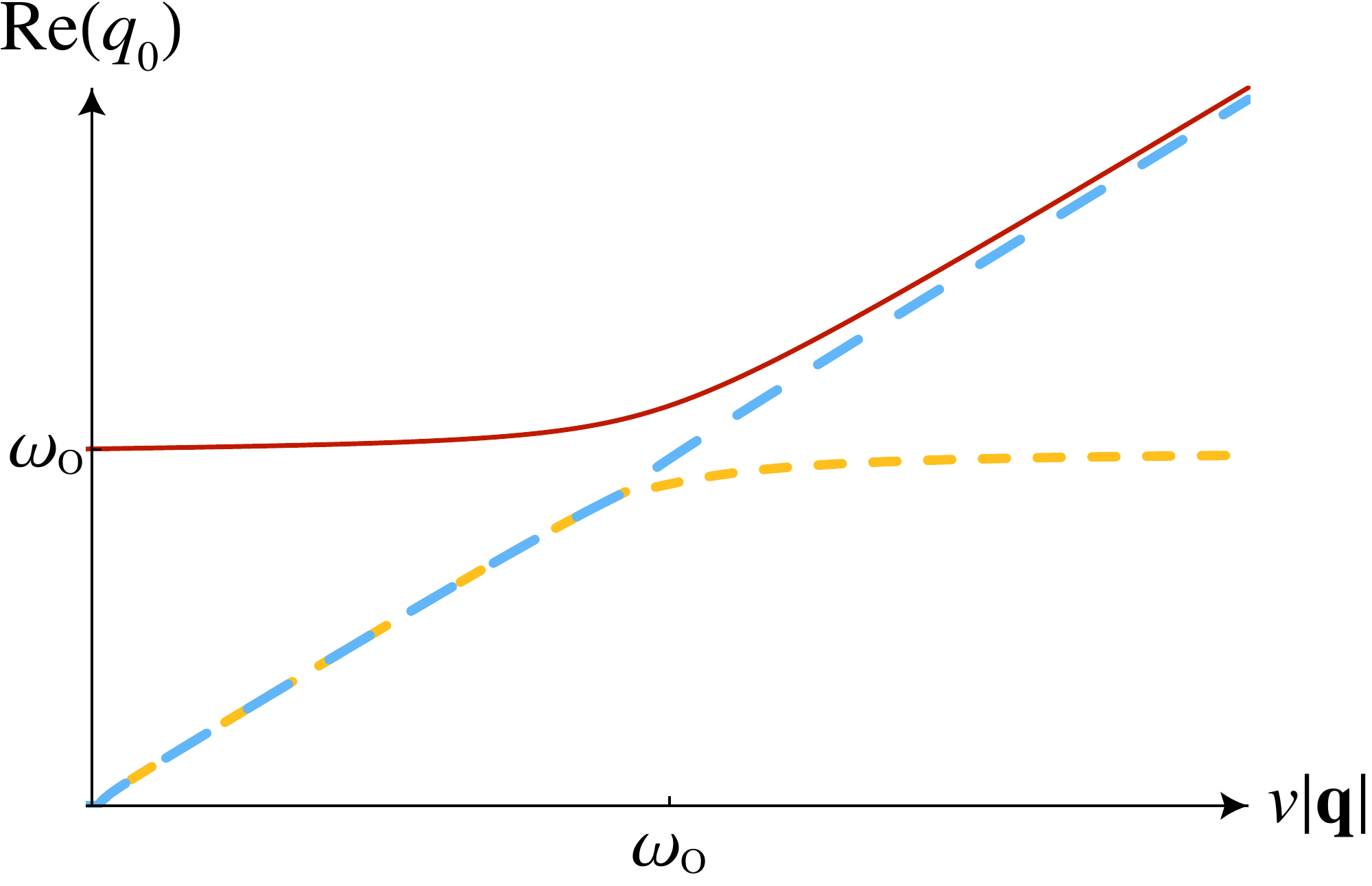}\llap{ \put(-120,0){\raisebox{0.7cm}{\fbox{\includegraphics[width=0.165\columnwidth]{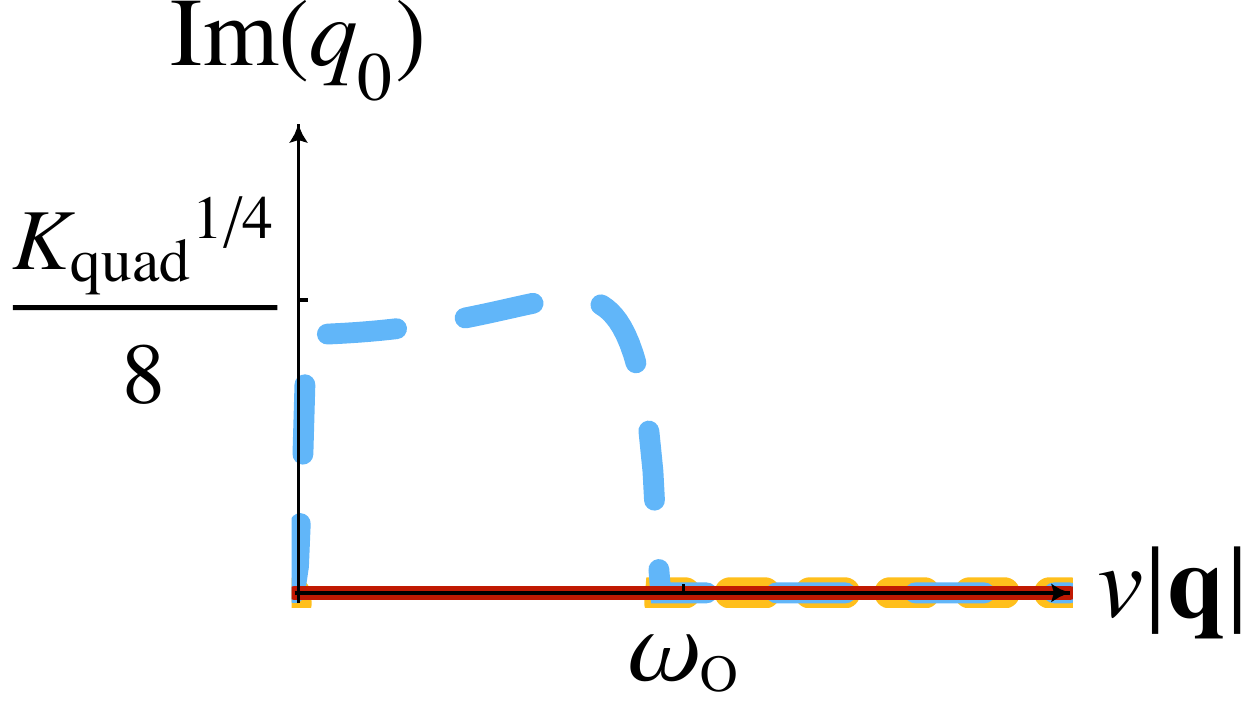}}}}}}\llap{\raisebox{4cm}{\shiftleft{7cm}{$\hat{\bf q}\cdot{\bf B}_0=-5\text{T}$}}}

}
\caption{Numerical solution of Eq.~(\ref{eq:eomnum}). From top to bottom and left to right we have the real and imaginary parts of the solutions for different values of  $\mathbf{\hat{q}}\cdot\mathbf{B}_{0}$, and $\hbar\omega_{0}=10\text{meV}$. The remaining parameters are described in the text. 
The magnetic field induces a hybridization between the ordinary optical phonon and the extraordinary quasilinear mode. The phonon linewidth at resonance ($|{\bf q}| =\omega_0/v$) depends strongly on the sign of $\mathbf{\hat{q}}\cdot\mathbf{B}_{0}$.
 When $|{\bf q}|\lesssim\omega_0/v$, the linear mode is doubly degenerate but one solution is unstable (anti-damped). In the inset, only the imaginary parts of stable modes are showed.}
\label{fig:manyT}
\end{figure}

Aside from the phonon dispersion, Eq.~(\ref{eq:edmfin}) enables to obtain the lattice contribution to the electric susceptibility.
Defining the lattice contribution to the polarization as
\begin{equation}
{\bf P}^{\rm latt} ({\bf q},q_0)=\frac{1}{\sqrt{N} {\cal V}_{\rm cell}}\sum_\lambda {\bf Q}_{{\bf q}\lambda}(q_0) v_{{\bf q}\lambda}(q_0),
\end{equation}
and extracting $v_{{\bf q}\lambda}$ from Eq.~(\ref{eq:edmfin}), we arrive at 
$P^{\rm latt}_i = \epsilon_0 \sum_j\chi^{\rm latt}_{ij} E_j$, where   
\begin{align}
\chi_{ij}^{\text{latt}}(\mathbf{q},q_{0}) & =\frac{1}{M {\cal V}_{\rm cell} \epsilon_0}\sum_{\lambda\lambda'}Q_{\mathbf{q}\lambda i}\left(\mathcal{D'}(\mathbf{q}, q_0)^{-1}\right)_{\lambda\lambda'}Q_{\mathbf{q}\lambda'j}^{*}\label{eq:SG}
\end{align}
is the lattice susceptibility, $i,j\in\{x,y,z\}$, and 
\begin{equation}
\mathcal{D'}_{\lambda\lambda'}({\bf q},q_0)  =\left[\left(\omega_{\mathbf{q}\lambda}^{2}-q_0^2\right)\delta_{\lambda\lambda'}-\frac{e {\cal V}}{M \hbar^2 \pi^2} \frac{g_{0z}^\lambda ({\bf q})\, g_{00}^{\lambda'}({\bf q})\,q_0\, {\bf q}\cdot{\bf B}_0}{q^2}\right].
\end{equation}
The lattice susceptibility can be measured e.g. via reflectivity experiments. 
In this case, $q_0=c |{\bf q}|$ and therefore the collective mode with $q^2\simeq 0$ is out of reach.
Instead, the main signature of the chiral anomaly consists of a magnetic-field-induced infrared activity for optical phonons that were intrinsically IR inactive (because  ${\bf Q}_\lambda \neq 0$ in spite of ${\bf Q}_\lambda^{(0)}=0$).

\section{Influence of the chiral anomaly in Raman scattering amplitude \label{app:appE}}

In this section, we will give details for the derivation of Eq. (16) from the main text.
Up to a numerical prefactor that depends on convention, the second-order Raman tensor is defined via~\cite{cardona1982} 
\begin{equation}
R_{i j}\propto \partial\chi_{i j}/\partial \xi_{{\bf q}\lambda},
\end{equation}
where $\xi_{{\bf q}\lambda}$ is related to the normal mode coordinate via $v_{{\bf q}\lambda}=\xi_{-{\bf q}\lambda}+\xi_{{\bf q}\lambda}^*$. 
From a diagramatic viewpoint, the susceptibility $\chi$ is a bubble diagram with two photon legs and an electron-hole pair.
Similarly, $\partial\chi/\partial \xi$, which describes the change in the {\em electronic} susceptibility due to lattice vibrations, can be ascribed to a triangle diagram with two photon legs and one phonon leg.
The fact that the triangle diagram analyzed in the main text corresponds to a time-ordered correlation function takes care of all the possible orderings in the scattering between the electron-hole pair, the photons and the phonon.
The phonon vertex contains an ordinary part as well as an axial part, the latter given by $c_5^\mu$. 
There is thus a direct link between a Feynman diagram containing the chiral anomaly and the Raman scattering amplitude measured in the laboratory.

Let us consider the case where no static magnetic fields are present.
The axial longitudinal part of the triangle diagram, including the external gauge fields as well as the internal phonon fields, reads
\begin{equation}
\label{eq:tri}
T^{(l)}_{\alpha\mu\nu} c_5^\alpha(q) A^\mu (k) A^\nu (k')=  c_5^\alpha(q) A^\mu (k) A^\nu (k') \frac{q_\alpha}{q^2}\epsilon_{\mu\nu\rho\sigma}k^\rho k'^\sigma=\frac{q_\alpha c_5^\alpha(q)}{8\pi^2 q^2} \left[{\bf E}(k)\cdot{\bf B}(k')+ k\leftrightarrow k'\right],
\end{equation}
where $k^\mu=-(\omega_{\rm in}, v {\bf k}_{\rm in})$, $k'^\mu=(\omega_{\rm out},v {\bf k}_{\rm out})$ are the momenta of the incoming and outgoing photons and $q^\mu=k^\mu+k'^\mu=(\omega_{{\bf q}\lambda}, v {\bf q})$ is the phonon momentum.
In a typical Raman scattering experiment, the incoming and outgoing photons are in the visible spectrum.
By Farady's law, we write
\begin{equation}
T^{(l)}_{\alpha\mu\nu} c_5^\alpha(q) A^\mu (k) A^\nu (k') = \frac{v}{8\pi^2} \frac{c_5^\alpha(q) q_\alpha}{q^2}\left[\frac{1}{k_0'} {\bf E}(k)\cdot[{\bf k}'\times {\bf E}(k')] + \frac{1}{k_0} {\bf E}\cdot[{\bf k}\times{\bf E}(k)]\right].
\end{equation}
Next, using ${\bf k}=-{\bf k}_{\rm in}$, ${\bf k}'={\bf k}_{\rm out}$, $k_0=-\omega_{\rm in}$, $k'_0=\omega_{\rm out}$, ${\bf E}(k)=E(k) \hat{\bf e}_{\rm in}$ and ${\bf E}(k')=E(k') \hat{\bf e}_{\rm out}$, we have
\begin{align}
T^{(l)}_{\alpha\mu\nu} c_5^\alpha(q) A^\mu (k) A^\nu (k') &=\frac{v}{8\pi^2} \frac{c_5^\alpha(q) q_\alpha}{q^2} \left[\frac{1}{k_0'} \hat{\bf e}_{\rm in}\cdot({\bf k}'\times\hat{\bf e}_{\rm out}) + \frac{1}{k_0} \hat{\bf e}_{\rm out}\cdot({\bf k}\times\hat{\bf e}_{\rm in})\right] E(k) E(k')\nonumber\\
&=\frac{v}{8\pi^2} \frac{c_5^\alpha(q) q_\alpha}{q^2} \left[\frac{1}{k_0} \hat{\bf e}_{\rm out}\cdot({\bf k}\times\hat{\bf e}_{\rm in})-\frac{1}{k_0'} \hat{\bf e}_{\rm out}\cdot({\bf k}'\times\hat{\bf e}_{\rm in})\right] E(k) E(k')\nonumber\\
&=\frac{v}{8\pi^2} \frac{c_5^\alpha(q) q_\alpha}{q^2} \left[\frac{1}{\omega_{\rm in}} \hat{\bf e}_{\rm out}\cdot({\bf k}_{\rm in}\times\hat{\bf e}_{\rm in})-\frac{1}{\omega_{\rm out}} \hat{\bf e}_{\rm out}\cdot({\bf k}_{\rm out}\times\hat{\bf e}_{\rm in})\right] E(k) E(k')\nonumber\\
&=\frac{v}{8\pi^2} \frac{c_5^\alpha(q) q_\alpha}{q^2} \frac{n}{c}\left[\hat{\bf e}_{\rm out}\cdot(\hat{\bf k}_{\rm in}\times\hat{\bf e}_{\rm in})- \hat{\bf e}_{\rm out}\cdot(\hat{\bf k}_{\rm out}\times\hat{\bf e}_{\rm in})\right] E(k) E(k'),
\end{align}
where $n$ is the refractive index of the Weyl semimetal.
Hence, the longitudinal part of the axial contribution to the Raman tensor for a given phonon mode $\lambda$ is given by
\begin{equation}
\label{eq:anram}
R^{\rm ax}_\lambda(\hat{\bf e}_{\rm in}, \hat{\bf e}_{\rm out}) \propto \frac{\partial^3\left[T^{(l)}_{\alpha\mu\nu} c_5^\alpha(q) A^\mu (k) A^\nu (k')\right]}{\partial {\bf E}(k) \partial {\bf E} (k') \partial v_{{\bf q}\lambda}(q_0)}
\propto \frac{q_\alpha}{q^2} \frac{\partial c_5^\alpha({\bf q}, q_0)}{\partial v_{{\bf q}\lambda}(q_0)}
\left(\hat{\bf k}_{\rm in}-\hat{\bf k}_{\rm out}\right)\cdot\left(\hat{\bf e}_{\rm in}\times\hat{\bf e}_{\rm out}\right)
\end{equation}
This contribution vanishes if $\hat{\bf e}_{\rm in}=\hat{\bf e}_{\rm out}$, and it is antisymmetric under $\hat{\bf e}_{\rm out}\leftrightarrow \hat{\bf e}_{\rm in}$.
According to Ref.~\cite{cardona1982}, if we approximate the incident and scattered radiation as having the same frequency (this is not a bad approximation for photons in the visible spectrum), then antisymmetric components in the Raman tensor can be introduced only by magnetic fields.
In other words, Raman tensors in non-magnetic semiconductors are expected to be symmetric.
At first sight, the antisymmetry of Eq.~(\ref{eq:anram}) in a time-reversal symmetric Weyl semimetal appears to be at odds with the preceding statement.
However, there is no contradiction because Eq.~(\ref{eq:anram}), which is proportional to $q_0$ in a time-reversal-symmetric Weyl semimetal, would vanish if we neglected the difference in frequency between the incident and the scattered light.
Another potentially interesting feature of Eq.~(\ref{eq:anram}) is that it predicts a resonance in Raman scattering when the phonon frequency is close to $v|{\bf q}|$ (i.e. when $q^2=0$).

Thus far, we have considered simply the contribution from the longitudinal part of the triangle diagram to the Raman tensor, and we have identified a potentially interesting resonance effect when $q^2=0$.
However, because $k'^2=\omega_{\rm out}^2-v^2 |{\bf k}_{\rm out}|^2$ and $k^2=\omega_{\rm in}^2-v^2 |{\bf k}_{\rm in}|^2$ are non null in Raman experiments, we anticipate (based on Ref.~[\onlinecite{armallis2009}]) that the $1/q^2$ pole in Eq.~(\ref{eq:anram}) will be cancelled by similar poles originating from the transverse part of the triangle diagram.
To check this, we begin by isolating all the terms that contain a $1/q^2$ pole in the triangle diagram:
\begin{equation}
\label{eq:sing}
\left[w_L-w_T^{(+)} \frac{k^2+k'^2-q^2}{q^2} + w_T^{(-)} \frac{k^2-k'^2}{q^2}\right] q_\alpha \epsilon_{\mu\nu\rho\sigma}k^\rho k'^\sigma,
\end{equation}
where $w_L=-4 i/q^2$ has the pole linked to the chiral anomaly.
Unlike $w_L$, the transverse form factors $w_T^{(\pm)}$ and $\tilde{w}^{(-)}_T$ are non singular at $q^2\to 0$, provided that $k^2\neq 0 \neq k'^2$.
This is why $\tilde{\omega}^T$ makes no appearance in Eq.~(\ref{eq:sing}).
However, we include $\omega_T^{(\pm)}$ in Eq.~(\ref{eq:sing}) because some of the kinematic factors that multiply them are singular at $q^2\to 0$.

Using Eq.~(\ref{eq:armi}), we Taylor expand Eq.~(\ref{eq:sing}) near $q^2=0$ and we obtain
\begin{equation}
\frac{i}{\pi^2 (k^2-k'^2)^3}\left[k^4 \ln\left(\frac{k^2}{q^2}\right) - k'^4 \ln\left(\frac{k'^2}{q^2}\right)\right] q_\alpha \epsilon_{\mu\nu\rho\sigma}k^\rho k'^\sigma +...(\text{terms non-singular in $q^2$})
\end{equation}
It is interesting that the $1/q^2$ pole from $w_L$ has been cancelled, as would have been expected from Ref.~[\onlinecite{armallis2009}].
This cancellation originates from the term $(s_1-s_2)/\sigma$ in the expression for $w_T^{(-)}$ (cf. Eq.~(\ref{eq:armi})).
However, a logarithmic singularity remains.
This singularity is accessible only to optical phonons (in the case of acoustic phonons, $\omega_{{\bf q}\lambda}=c_\lambda |{\bf q}|$, where the sound velocity is much smaller than the Fermi velocity, hence preventing $q^2=0$).

The contribution of the singular terms to the Raman tensor is
\begin{equation}
\label{eq:anram2}
R^{\rm ax}_\lambda(\hat{\bf e}_{\rm in}, \hat{\bf e}_{\rm out}) \propto q_\alpha\frac{\partial c_5^\alpha({\bf q}, q_0)}{\partial v_{{\bf q}\lambda}(q_0)}
\frac{1}{(k^2-k'^2)^3}\left[k^4 \ln\left(\frac{k^2}{q^2}\right) - k'^4 \ln\left(\frac{k'^2}{q^2}\right)\right] \left(\hat{\bf k}_{\rm in}-\hat{\bf k}_{\rm out}\right)\cdot\left(\hat{\bf e}_{\rm in}\times\hat{\bf e}_{\rm out}\right),
\end{equation}
which agrees with Eq. (16) of the main text (note that we adopt a slightly different notation therein).
In sum, we have identified a potentially interesting logarithmic singularity in the Raman tensor of Weyl semimetals.
Since this singularity is independent from the ultraviolet cut-off of the theory, it is associated to low-energy universal properties of 3D massless Dirac fermions.
Although resonance effects in Raman scattering have been widely documented, the ``infrared'' resonance at $q^2=0$ may be considered a characteristic signature of Weyl fermions.

\section{Robustness of the anomaly pole under a finite ultraviolet cutoff \label{app:appF}}

In the preceding sections, we have borrowed results from high-energy physics in order to compute the triangle diagram and the corresponding $1/q^2$ anomaly pole.
In the derivation of the high-energy results, the ultraviolet cutoff of the theory is taken to be infinite.
Yet, in condensed matter, the Weyl Hamiltonian with linear dispersion is valid only at low energies, below a certain finite cutoff $\Lambda$.
Hence, one may wonder whether the anomaly pole will need to be corrected from finite $\Lambda$ effects.
In this section, we confirm that the $1/q^2$ pole in the longitudinal part of the triangle diagram is robust under finite $\Lambda$ effects, provided that the photon momenta appearing in the triangle diagram are small compared to $\Lambda$.

For simplicity , we will assume zero temperature and chemical potential, and we will consider the triangle amplitude $T_{\alpha\mu\nu}(k,k')$ in a form that is slightly different from (but equivalent to it through a momentum shift) Eq.~(\ref{eq:Gamma}):
\begin{align}
T_{\alpha\mu\nu}(k,k') & =\int_{p}\text{Tr}\left[\frac{1}{\cancel{p}-\cancel{l}}\gamma^{5}\gamma_{\alpha}\frac{1}{\cancel{p}+\cancel{l}}\gamma_{\mu}\frac{1}{\cancel{p}-\cancel{r}}\gamma_{\nu}\right]\label{eq:AppT-1}\\
\text{with} & \begin{cases}
l=\frac{k+k'}{2}=\frac{q}{2}\\
r=\frac{k-k'}{2}
\end{cases}
\end{align}
and $\cancel{p}\equiv \gamma^\lambda p_\lambda$.
Hereafter, we will concentrate on the longitudinal part of $T_{\alpha\mu\nu}$, and we will determine the fate of its $1/q^2$ when the ultraviolet cutoff is finite.  

Using $\left(\cancel{p}-\cancel{l}\right)^{-1}2\cancel{l}\left(\cancel{p}+\cancel{l}\right)^{-1}=\left(\cancel{p}-\cancel{l}\right)^{-1}-\left(\cancel{p}+\cancel{l}\right)^{-1}$
as well as the invariance of $T_{\alpha\mu\nu}$ under the exchange
of indices $\mu$ and $\nu$, and computing the traces of Dirac matrices,
we get 
\begin{align}
q_\alpha T^{\alpha\mu\nu}(k,k') & =4\text{i}\lim_{\Lambda\rightarrow\infty}\intop_{p<\Lambda}\left[f_{1}(p)-f_{1}(p+k)\right]+\left[f_{2}(p)-f_{2}(p+k')\right]\label{eq:axdiv1-1}\\
f_{1}(p) & =\epsilon^{\sigma\rho\mu\nu}\frac{k_{\sigma}k_{\rho}^{'}-2p_{\sigma}k_{\rho}^{'}}{\left(p-k-k'\right)^{2}\left(p-k+k'\right)^{2}}\\
f_{2}(p) & =\epsilon^{\sigma\rho\mu\nu}\frac{k_{\sigma}^{'}k_{\rho}-2p_{\sigma}k_{\rho}}{\left(p-k-k'\right)^{2}\left(p+k-k'\right)^{2}}
\end{align}
Assuming that the photon momenta $k$ and $k'$ are small compared to $\Lambda$,  we may expand
\begin{align}
\intop_{p<\Lambda}f_{1}(p+k)-f_{1}(p) & =\intop_{p<\Lambda}\left[k^{\beta}\partial_{p_{\beta}}f_{1}(p)+\frac{1}{2}k^{\beta}k^{\eta}\partial_{p_{\eta}}\partial_{p_{\beta}}f_{1}(p)+\frac{1}{6}k^{\beta}k^{\eta}k^{\gamma}\partial_{p_{\gamma}}\partial_{p_{\eta}}\partial_{p_{\beta}}f_{1}(p)\right]+\mathcal{O}(k^{4})\\
 & =\frac{\Lambda^{2}}{8\pi^{2}}\left[k^{\beta}\left\langle \Lambda_{\beta}f_{1}(\Lambda)\right\rangle +\frac{1}{2}k^{\beta}k^{\eta}\left\langle \Lambda_{\beta}\partial_{\eta}f_{1}(\Lambda)\right\rangle +\frac{1}{6}k^{\beta}k^{\eta}k{}^{\gamma}\left\langle \Lambda_{\beta}\partial_{\gamma}\partial_{\eta}f_{1}(\Lambda)\right\rangle \right]+\mathcal{O}(k^{4}),\label{eq:exp1-1}
\end{align}
where $\left\langle \:\right\rangle $ denotes an average
over a 3D hyperspherical surface at $p=\Lambda$. 
One may proceed in the same fashion with the terms involving $f_2$.
At large $\Lambda$, $f_1(\Lambda)$ and $f_2(\Lambda)$ scale as $1/\Lambda^3$, and every derivative acting on them brings about an additional factor of $1/\Lambda$. 
After a straightforward expansion of $f_{1}$ and $f_{2}$ as well as their derivatives, and using
\begin{align*}
\left\langle \Lambda_{i}\right\rangle  & =\left\langle \Lambda_{i}\Lambda_{j}\Lambda_{k}\right\rangle =0\\
\left\langle \Lambda_{i}\Lambda_{j}\right\rangle  & =\frac{1}{2\pi^{2}}\int\text{d}^{3}\Omega\Lambda_{i}\Lambda_{j}=\frac{1}{4}g_{ij}\Lambda^{2}\\
\left\langle \Lambda_{i}\Lambda_{j}\Lambda_{k}\Lambda_{l}\right\rangle  & =\frac{1}{2\pi^{2}}\int\text{d}^{3}\Omega\Lambda_{i}\Lambda_{j}\Lambda_{k}\Lambda_{l}=\frac{1}{24}\left(g_{ij}g_{kl}+g_{ik}g_{jl}+g_{il}g_{jk}\right)\Lambda^{4},
\end{align*}
 we get
\begin{equation}
q_{\alpha}T^{\alpha\mu\nu}(k,k')=\frac{\text{i}\epsilon^{\sigma\rho\mu\nu}}{2\pi^{2}}\left[1-\frac{9}{4}\frac{k^{2}+k'^{2}}{\Lambda^{2}}+\mathcal{O}(\frac{1}{\Lambda^{3}})\right]k_{\sigma}k_{\rho}^{'}.\label{eq:Tcorrec-1}
\end{equation}
This equation implies that $T^{\left(l\right)}_{\alpha\mu\nu}\propto q_\alpha/q^2$, i.e. a pole at $q^2=0$, even when $\Lambda$ is finite. 
Moreover, the coefficient of this pole has a dominant cutoff-independent part when the photon momenta are small compared to the cutoff. 

\end{widetext}


\begin{thebibliography}{99}
\bibitem{cutoff} In real WSM, the frequency of visible photons may be larger than the typical energy scale in which the electronic dispersion can be approximated as linear.
Hence, we expect that Eq.~(\ref{eq:anram2}) will have corrections from high-energy electronic states.
Nevertheless, the $\ln(q^2)$ dependence should be robust because it has topological origin.
\end{thebibliography}
\end{document}